\documentstyle[12pt,epsf,psfig]{article}

\def\beq{\begin{equation}}
\def\eeq{\end{equation}}
\def\beqa{\begin{eqnarray}}
\def\eeqa{\end{eqnarray}}

\def\a {{\rm f}}
\def\T{T}
\def\U{U}
\def\gi{{\cal{G}}_{(f_i)}(\nu_i)}
\def\e{r}
\def\X{X}
\def\Y{Y}
\def\blocA{\Gamma_{3 \times 3}}
\def\blocB{\Gamma_{6 \times 6}}

\begin{document}

\begin{flushright}
ITP-SB-98-23\\
EDINBURGH 98/3
\end{flushright}

\begin{center}{\bf\Large\sc Evolution of Color Exchange}
\vglue 0.25cm
{\bf\Large\sc  In QCD Hard Scattering}
\vglue 1.2cm
\begin{sc}
Nikolaos Kidonakis\\
\vglue 0.5cm
\end{sc}
{\it Department of Physics and Astronomy\\
University of Edinburgh,
Edinburgh EH9 3JZ, Scotland, UK}\\
\vskip 1.0cm
\begin{sc}
Gianluca Oderda and George Sterman\\
\vglue 0.5cm
\end{sc}
{\it Institute for Theoretical Physics \\
SUNY at Stony Brook,
Stony Brook, NY 11794-3840, USA}\\
\end{center}
\vglue 1cm

\begin{abstract}
In QCD hard scattering cross sections, the color content of the underlying hard 
scattering evolves with a factorization scale.
This evolution is controlled by an
anomalous dimension matrix, specific to each hard-scattering
reaction.  Anomalous dimensions are determined from the renormalization
of products of ordered exponentials of the gauge field,
which describe the coherent radiation of gluons by incoming
hadrons and the observed jets or particles of the final state.
The anomalous dimensions depend on the kinematics of the 
underlying hard scattering, but are free of collinear singularities.
A number of these matrices are available in the literature.  Here,
we exhibit one-loop mixing matrices  for the full list of $2\rightarrow 2$
reactions involving light quarks and gluons.  The
eight-by-eight anomalous dimension matrix for gluon-gluon
scattering shows a simplified structure in the basis 
corresponding to definite color exchange in the $t$-channel.
\end{abstract}

\pagebreak

\section{Introduction}

Inclusive, short-distance hadronic cross sections
may be factorized into convolutions of universal, nonperturbative
parton distributions and/or fragmentation functions,
and perturbative hard-scattering coefficient functions.
In this procedure, a factorization scale $\mu$
separates nominally long-
and short-distance components of the cross section.
When $\mu$ is chosen large enough that
the coupling $\alpha_s(\mu^2)$ is perturbative,
it is  possible to calculate the
coefficient functions, and hence to determine
the cross section, assuming that the relevant nonperturbative
densities are known.  

Factorization is an expression of quantum mechanical incoherence.
The parton distributions summarize the internal dynamics
of the incoming hadrons, and of their fragments that
evolve into the final state.  These distributions are
independent of each other and of the fragmentation of
scattered partons into jets and hadrons, in the final state.  They
are also incoherent with the short-distance dynamics
that produces the hard scattering.  Finally, they are
incoherent with the dynamics of ``noncollinear"
gluons of any energy -- from the hard scale $Q$ down to zero --
that are radiated away from the incoming hadrons, or from
the observed components of the final state \cite{CSS1,CSS88,CSSrv}.

Each of these components of a hard-scattering process:
parton distributions, parton (or jet) fragmentation,
hard scattering and soft gluons, may be thought of as
governed by an effective field theory, in which the parton dynamics
at differing scales and directions has been removed,
and sometimes replaced by specific operators.  At fixed order, 
coefficient functions combine contributions of quanta that are
off-shell by the order of $Q$ with those from noncollinear soft gluons,
because divergent soft-gluon behavior cancels, leaving
finite contributions at each order in perturbation theory.
These contributions, however, retain sensitivity
to momentum scales much smaller than $\mu$, and may
produce large, although finite, higher-order corrections.

Noncollinear soft-gluon dynamics is an important ingredient 
in a variety of resummed cross sections.
Their first complete treatment was given by Collins and Soper
for the transverse momentum ($Q_T$) distribution of hadron pairs 
in ${\rm e}^+{\rm e}^-$ annihilation \cite{CoSo81}.
Because this cross section is not fully inclusive, it is
necessary to resum (Sudakov) logarithms of $Q_T$ due to soft
gluon emission to determine the cross section near $Q_T=0$,
a procedure intensively studied for Drell-Yan pairs \cite{DYqt1,DYqt2}.  
As is typical of such resummations, nonperturbative effects
remain important phenomenologically at realistic energy scales.  In fact, the perturbative resummation
suggests the functional form of these effects \cite{DYqt2}.  A similar
resummation is necessary to define elastic scattering amplitudes
for hadrons and partons at large momentum transfer and 
fixed angle \cite{BottsSt,SotiSt,GK,KK}.  In both
of these cases, leading logarithms are generated by  
soft gluons emitted collinear to incoming
partons.  Nevertheless, a treatment of noncollinear soft emission,
{\it i.e.} soft gluons emitted at angles between the directions
of the primary incoming and outgoing hard partons, is
necessary to extend the formalism to next-to-leading logarithm
and beyond.  The pattern of this emission is essentially
coherent \cite{CoCoh}; as we shall see below, it is 
determined primarily by interference between emission
from different hard partons.
Despite this, it remains incoherent with the collinear dynamics of incoming
and outgoing hard partons. That is, it factorizes from parton distributions
and from suitably defined final-state jets.

In inclusive quantities such as the Drell-Yan cross section
$d\sigma/dQ^2$, soft gluon dynamics  is somewhat hidden in
higher order corrections to short-distance coefficient functions.
It appears indirectly, through integrable logarithms at the ``edge of phase space",
where the observed final state uses up all partonic energy, leaving
none over for  QCD radiation.  These phenomena have been
 studied in inclusive electroweak reactions such as the Drell-Yan
cross section \cite{St87,CT},  
and more recently in inclusive QCD reactions such as
heavy quark \cite{LSvNKV,BC,CMNT,KS,Thesis,NKJSRV,BCMN} 
and dijet production \cite{KOS1}.  

Beyond leading logarithms, all of these resummations
require a treatment of soft, noncollinear, gluons.   The hard
scattering of partons of specific flavors involves, in
general, a number of possible color exchanges.  On quite
general grounds, we expect this information to be reflected
in observable properties of the final state \cite{Color_eff}.  
At the same time, it is clear that if soft gluons
are factorized from the hard scattering, the color content
of the latter depends on the scale at which the
separation is carried out.  Nevertheless, physical
cross sections cannot depend on this scale.  The combination of factorization, and
invariance under scale choice ensures the possibility of
resummation in the scale dependence \cite{CLS}.  

In this procedure, the emission of noncollinear
soft gluons is described by 
matrix elements of products of 
ordered exponentials of the gluon field, coupled at a point
representing the hard scattering \cite{BottsSt,KS,Thesis,BCMN,KOS1}.  
The factorization scale between
the hard and soft gluon functions may be interpreted as the
renormalization scale for these operators.  If the physical
cross section is defined so that the soft gluon function depends
on a single ratio of a phenomenological mass scale to the
renormalization scale, then the entire soft function is determined
by its  anomalous dimensions.  

Rather than review the specifics
of QCD resummations in detail, we shall treat here one feature common to them
all, the evolution of color exchange in noncollinear soft gluon
emission.  Our discussion will therefore treat gluons interacting
with ordered exponentials instead of physical partons.  We will
present the full set of anomalous dimensions relevant to the
scattering of light partons, including quark-quark, quark-antiquark
(which have been presented previously), quark-gluon, and (by far the
most complex case) gluon-gluon.  Each of these partons
will be replaced, for this discussion, by Wilson lines of the
corresponding color representation.  The justification for
this approximation, and the detailed role that it plays in each
cross section, may be found in the references cited above.

We begin, in Sec.\ 2,
with a discussion of the renormalization of products
of ordered exponentials, with the aim of isolating the
evolution of color mixing.  We shall see that, although
the analysis immediately encounters collinear divergences,
it is easy to identify finite anomalous dimensions that
govern the mixing of color.   We go on, in Sec.\ 3, to recall, for definiteness,
the soft gluon contribution to 
inclusive jet cross sections in hadronic scattering \cite{KOS1}.  This
``soft function" is manifestly free of collinear singularities, and 
its evolution, with mixing, reflects the color content of the underlying
hard scattering.  
In Sec.\ 4, we introduce a color basis notation
that is convenient for the solution of the evolution equation
for the soft function.  This is followed, in Secs.\ 5 and 6, by the computation of
the anomalous dimension matrices for various processes, along with
their eigenvalues and eigenvectors.  We have included two
appendices that describe some of the details of the calculations. 

\section{Soft Anomalous Dimensions from Wilson Lines}

Our discussion begins
with ordered exponentials, or Wilson lines.
A general Wilson line is labelled by a path
in space-time, $C$, beginning at point $z$ 
and ending at point $z'$,
\beq
W[C;z',z]= P\exp \left[ -ig\int_{\lambda_1}^{\lambda_2} d\eta\, {dy (\lambda)\over d\eta}\cdot
A\left(y(\eta) \right) \right ] \, .
\label{genline}
\eeq
The variable of
integration $\eta$  parameterizes 
$C$, with $y(\lambda_1)=z$, $y(\lambda_2)=z'$.

Wilson lines have myriad applications in quantum field
theory, but their interest here is as
a model for the radiation of soft gluons by  fast-moving 
partons (quarks and gluons).   Any hard-scattering event, be it
based on electroweak or QCD hard scattering, results in nonabelian radiation, in 
much the same manner as in classical electrodynamics.  Indeed,
as in classical electrodynamics, the pattern of soft radiation
is determined by the charge currents
at long times before and after the scattering
event.  The factorization properties of inclusive hard-scattering
cross sections in QCD rely on this feature \cite{CSSrv}.  

Initial and final-state partons are
represented by ordered exponentials in
which $C$ is a straight line, in the direction 
of the four-velocity, $\beta$, of the relevant parton.  
For inclusive cross sections, the lines
extend from a common point
$x$ to infinity, either from the distant past, or toward the distant future.
We shall adopt the notation of Ref.\ \cite{KOS1} for such Wilson
lines,
\beq
\Phi_{\beta}^{(f)}({\lambda}_2,{\lambda}_1;x)
=
P\exp\left(-ig\int_{{\lambda}_1}^{{\lambda}_2}d{\eta}\; 
{\beta}{\cdot} A^{(f)} ({\eta}{\beta}+x)\right)\, ,
\label{eq:wilson} 
\eeq
where  the gauge field $A^{(f)}$ is a matrix
in the representation of the flavor $f$.
Comparing this definition with
Eq.\ (\ref{genline}), we have $y(\eta)={\eta}{\beta}+x$.

Although the Wilson lines $\Phi$ in Eq.\ (\ref{eq:wilson})
extend to infinity, they reproduce the dynamics of the
gauge field for only a finite time before and after
a hard scattering.  At long times, nonperturbative
dynamics dominate, nonabelian charges are neutralized,
and a picture in terms of isolated currents fails utterly.
Therefore, the procedure applies only to
cross sections for which such long-distance effects
{\it cancel} in perturbation theory.  The final answers for such quantities
are insensitive to the infrared
properties of amplitudes generated from Wilson lines
of the form of Eq.\ (\ref{eq:wilson}), although
it may be necessary to deal with these properties at
intermediate steps in any calculation.  In fact,
it is their ultraviolet behavior that we shall find most
interesting.  

This approach has been applied to
the classic electroweak hard scattering processes,
${\rm e}^+{\rm e}^-$ annihilation,  DIS and Drell-Yan
vector boson production \cite{St87,CT,KM}. In each of these cases, the relevant
operator is a product of two $\Phi$'s coupled
at a color singlet vertex.  Equivalently it is a single Wilson line,
whose direction changes discontinuously at a single point.
For ${\rm e}^+{\rm e}^-$ annihilation, the 
curve $C$ consists of two straight paths, both
extending to positive infinity in time,  describing
the creation of a quark pair at a point.  For DIS, $C$ comes
in from negative infinity and changes direction, 
describing the scattering of  a quark or antiquark, while for Drell-Yan,
$C$ represents 
the annihilation of a pair.

\begin{figure}
\centerline{\epsffile{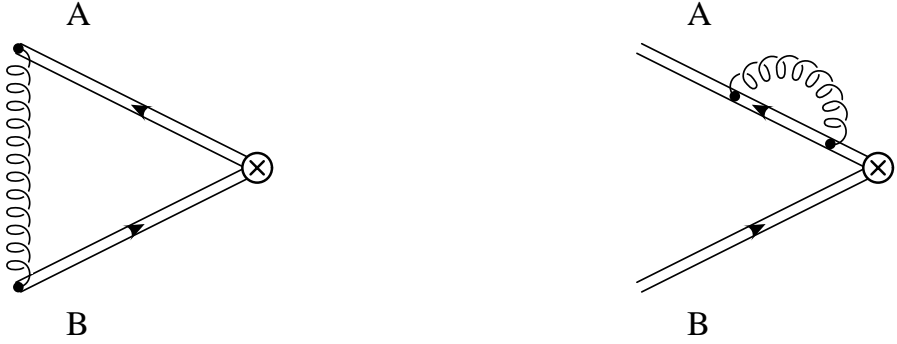}}
\caption{Eikonal vertex correction and eikonal self-energy for
the Drell Yan process.}
\label{eikoDY}
\end{figure}

For definiteness, we consider the Wilson line analog of 
the Drell-Yan cross section \cite{KM}, where
the configuration just described corresponds to the product
\begin{equation}
W^{\rm (DY)}_{b_2,b_1}(x)
=
\delta_{a_1a_2}\; 
\Phi_{\beta_2}^{(\bar{q})}(0,-\infty;x)_{a_2b_2}
\Phi_{\beta_1}^{(q)}(0,-\infty;x)_{a_1b_1}\, .
\label{WDY}
\end{equation}
The vertex at $x$ is associated with ultraviolet divergences,
as are self-energies of the eikonal lines.  The divergent one-loop
diagrams are shown in Fig.\ \ref{eikoDY}.   
These vertex and self-energy diagrams are conveniently renormalized by 
perturbative counterterms, corresponding to
multiplicative renormalization of the composite operators
\cite{Brandtetal,eikrenorm1,KR},
\beq
W^{\rm (DY)(0)}(x)=
Z_W^{\rm (DY)} \left(\alpha_s,{\beta_1\cdot\beta_2 \over \sqrt{\beta_1^2\beta_2^2}}\right) 
W^{\rm (DY)}(x)\, ,
\label{Wrenorm}
\eeq
where $W^{\rm (DY)(0)}$ is the bare operator, and
where we have used the invariance of $W$ under rescalings
of the velocities $\beta_i$.  In the following, we suppress
dependence on $x$, the position of the composite vertex.
The independence of the unrenormalized operator from
the renormalization scale $\mu$ implies that
the renormalized operator satisfies the renormalization group (RG) 
equation\footnote{Strictly speaking, the multiplicative renormalization in Eq.\ (\ref{Wrenorm}),
and the corresponding RG equation (\ref{RGW}) apply to the
renormalization of the composite operator only; the 
renormalization associated with the QCD Lagrangian is assumed
to have been carried out.  In {\protect \cite{Brandtetal}} the
renormalization proof was formulated in terms of the vacuum expectation
value of the corresponding loop.},
\beq
\mu{d\over d\mu}\; W^{\rm (DY)}
=\Gamma_W\left(\alpha_s,{\beta_1\cdot\beta_2 \over \sqrt{\beta_1^2\beta_2^2}}\right)\; 
W^{\rm (DY)}\, .
\label{RGW}
\eeq
The anomalous dimension $\Gamma_W(\alpha_s,\beta_i)$ is the same for all
color singlet products of the form of Eq.\ (\ref{WDY}),
and is sometimes referred to as $\Gamma_{\rm cusp}$ \cite{KR}.  

Because $W^{\rm (DY)}(\alpha_s,\beta_i,n)$ is a gauge-invariant operator, so
is $\Gamma_W$.  
A direct calculation verifies that, as suggested
by Eq.\ (\ref{RGW}), $\Gamma_W$ is somewhat
unusual for an anomalous dimension, in that it diverges
in the limit of light-like velocities for the external
lines, $\beta_i^2=0$.  Fixing the $\beta_i^2$ to be
small, but nonvanishing, we find the explicit expression
\beq
\Gamma_W={\alpha_s\over\pi}C_F\; \left[
\ln\left(-2\beta_1\cdot\beta_2\over \sqrt{\beta_1^2\beta_2^2}\right)
-1 \right]\, .
\label{GamWoneloop}
\eeq
The divergence at light-like velocities is a
collinear enhancement, due to the emission of 
gluons in the direction of the eikonal line.
In suitably-defined
inclusive cross sections, the corresponding collinear divergences either cancel,
or are absorbed into parton distributions \cite{CSSrv}. 
We shall see that the same is true for a product of two Wilson lines
like Eq.\ (\ref{WDY}) and its generalizations.  Specifically, the
collinear divergences in $\Gamma_W$ are universal,
and may be factored from the underlying local coupling.
In computing loop corrections to $W$ and $\Phi$,
we treat the eikonal lines as on-shell particles,
normalized by the square root of the residue of the eikonal two-point
function. This ensures gauge invariance.

To demonstrate factorizability,
we shall find it useful to work in an axial gauge, $n\cdot A=0$, with
$n\cdot \beta_i\ne 0$, $i=1,2$.  
In any gauge, a single Wilson line $\Phi_\beta^{(q)}$
in quark representation,
defined on a ray in the direction of velocity $\beta$ is
also multiplicatively renormalizable \cite{Brandtetal},
\beq
\Phi^{(q)(0)}_{\beta}=
Z_\Phi\left(\alpha_s,{| \beta\cdot n | \over \sqrt{\beta^2 |n^2|}}\right) \Phi^{(q)}_\beta\, ,
\label{Phirenorm}
\eeq
where again, rescaling invariance in the velocity and the gauge
vector limits possible arguments of $Z$.
Eq.\ (\ref{Phirenorm}) implies that $\Phi^{(q)}_\beta$
satisfies a renormalization group equation
\beq
\mu{d\over d\mu}\Phi^{(q)}_\beta
=\Gamma_\Phi^{(q)}\left(\alpha_s,{| \beta\cdot n | \over \sqrt{\beta^2 |n^2|}}\right)\; 
\Phi^{(q)}_\beta\, ,
\label{RGPhi}
\eeq
with $\Gamma_\Phi^{(q)}$ another anomalous dimension.
Clearly,
$\Phi^{(\bar{q})}$ requires the same renormalization constant.

The one-loop expression for   
 $\Gamma_\Phi$ in axial gauges has the same collinear ($\beta^2\rightarrow 0$) 
divergence as $\Gamma_W$, but now
through gauge-dependent logarithms,
\beq
\Gamma_\Phi^{(q)} \left(\alpha_s,{| \beta\cdot n| \over \sqrt{\beta^2 |n^2|}}\right)
={\alpha_s\over\pi}C_F\; \left[
\ln\left(2 | \beta\cdot n | \over \sqrt{\beta^2 | n^2 |}\right)
-1\right]\, .
\eeq
Evidently, the collinear divergences in this anomalous dimension
exactly match those of $\Gamma_W^{({\rm DY})}$, with the opposite sign.
As we shall see, this result is quite general.

To make use of this correspondence, let us define a new operator,
which generates the sum of  diagrams, containing the
Drell-Yan vertex, that cannot be reduced by cutting a single eikonal
line.  We do this by simply dividing out the vacuum expectation values
of the two external Wilson lines in Eq.\ (\ref{WDY}),
which defines $W^{({\rm DY})}$,
\beq
S^{({\rm DY})}=W^{\rm (DY)}\; \langle 0|\Phi_{\beta_1}^{(q)}(0,-\infty;0)|0\rangle^{-1}\;
\langle 0|\Phi_{\beta_2}^{(\bar{q})}(0,-\infty;0)|0\rangle^{-1}\, ,
\label{SDY}
\eeq
where in the vacuum expectation value a trace and an average of color of
the operator $\Phi$ is assumed.  

Given the RG equations for $W^{\rm (DY)}$ and the $\Phi$'s,
the function $S^{\rm (DY)}$ satisfies a renormalization group
equation of exactly the same sort as $W^{\rm (DY)}$, 
\beq
\mu{d\over d\mu}S^{\rm (DY)}
=\Gamma_S\left(\alpha_s,\beta_i, n \right)\; 
S^{\rm (DY)}\, ,
\label{RGS}
\eeq
with a
 ``soft" anomalous dimension,
\beqa
\Gamma_S&=&\Gamma_W-
\Gamma_\Phi^{(q)} \left(\alpha_s,{| \beta_1\cdot n| \over \sqrt{\beta_1^2 |n^2|}}\right)
 -\Gamma_\Phi^{(\bar{q})} \left(\alpha_s,{|\beta_2\cdot n| \over \sqrt{\beta_2^2 |n^2|}}\right)\nonumber\\
&=& {\alpha_s\over\pi}C_F\; \left[1+
\ln\left({-\beta_1\cdot \beta_2\; |n^2| \over 2 |\beta_1\cdot n| \, |\beta_2\cdot n|}\right)\right]\, .
\label{GammaSDYdef}
\eeqa
We may think of $\Gamma_S$ as the ``coherent" part of the
``cusp" anomalous dimension $\Gamma_W$, that is, the part
that depends on the specific
color-singlet vertex in Eq.\ (\ref{WDY}),
but is independent of the mutually incoherent collinear behavior of the lines
themselves. 
Our definition of $\Gamma_S$ is, however, somewhat arbitrary, and we might
change it through a  finite renormalization
and/or a change of gauge.  A natural choice is $A^0=0$ gauge
in a frame where $\beta_1^\mu=\delta_{\mu+}$ and 
$\beta_2^\mu=\delta_{\mu-}$,
in which case $\Gamma_S=(\alpha_s/\pi)C_F\left(1-i \pi \right)$ \cite{St87}, but we are
even free to make $\Gamma_S^{\rm (DY)}$ vanish.   
Any choice leads to the same 
amplitude.

The value of this approach is found by generalizing
$W^{\rm (DY)}$ to products of multiple 
Wilson lines.  For QCD hard-scattering processes,
products of four lines are most interesting,
and illustrate the general case.
As noted above, the collinear divergences 
represented by Eq.\ (\ref{GamWoneloop})
are more general than the color-singlet
vertex in $W^{(DY)}$, precisely
because they may be factored from this
vertex, and from each other.  In the anomalous
dimensions, which appear eventually
in exponents, this factorization is guaranteed 
to be additive. 

In the context of QCD hard-scattering \cite{KS,Thesis,NKJSRV,BCMN,KOS1}, 
most of the above considerations apply, although now with a 
more complex and interesting color flow.  In this paper, we
shall be concerned specifically with soft
radiation for cross sections due to $2\rightarrow 2$
scattering of light quarks and gluons at short distances,
of relevance to high-$p_T$ jet production.  Analogous reactions are
also at the basis of heavy quark production.  In this larger context,
the value of isolating single-logarithmic soft gluon
dynamics from corresponding, universal, collinear dynamics is clear. 
We shall review very briefly in Sec.\ 3 the role of products
of Wilson lines in hard cross sections, using as an example the resummation of
threshold corrections in jet cross sections.  First, however,
we show how the generalization of the color-singlet product
$W^{\rm (DY)}$ leads naturally to a matrix of anomalous dimensions
that describe the mixing of color with changing scales.

Following Ref.\ \cite{KOS1}, we introduce a class of operators 
coupling four Wilson lines, corresponding to the 
primary partons involved in a hard scattering,
\beqa
W_I^{(\a)}(x)_{\{\e_k\}}
&=& 
\sum_{\{d_i\}}
\Phi_{\beta_2}^{(f_2)}(\infty,0;x)_{\e_2,d_2}\; 
\Phi_{\beta_1}^{(f_1)}(\infty,0;x)_{\e_1,d_1}\cr
&\ &\times
\left( c_I^{(\a)}\right)_{d_2d_1,d_Bd_A}\; 
\Phi_{\beta_A}^{(f_A)}(0,-\infty;x)_{d_A,\e_A}
\Phi_{\beta_B}^{(f_B)}(0,-\infty;x)_{d_B,\e_B}\, ,
\nonumber\\
&\ &
\label{eq:wivertex}
\eeqa
and a corresponding irreducible soft function,
\beq
S_I^{(\a)}=W_I^{(\a)}\; 
\prod_{i=A,B}\langle 0|\Phi^{(f_i)}_{\beta_i}(0,-\infty;0)|0\rangle^{-1}\;
\prod_{i=1,2}\langle 0|\Phi^{(f_i)}_{\beta_i}(\infty,0;0)|0\rangle^{-1}\, .
\label{DYirredsoft} 
\eeq
As above, $\beta_i$ represents the
 four-velocities of initial-
and final-state partons.  The superscript $\a$ represents a 
scattering process, 
\beq
f_A(\ell_A,\e_A)+f_B(\ell_B,\e_B)\rightarrow f_1(p_1,\e_1)+f_2(p_2,\e_2)\, ,
\label{sctgprocess}
\eeq
involving four partons, with flavors $f_i$, incoming momenta $l_i$
and outgoing momenta $p_i$, and
color $\e_i$.
The color tensor $\left(c_I^{(\a)}\right)_{d_2d_1,d_Bd_A}$ describes 
the couplings of the ordered exponentials
with each other in color space.  If, for example, the incoming
lines are quark and antiquark, the indices $\e_i,d_j$ are
in the fundamental representations of $SU(3)$.
Examples of bases for
the color tensors $c_I$ then include:  singlet and octet matrices
in the $s$- or in the $t$-channel, or $s$-channel color singlet and
 $t$-channel color singlet.  

\begin{figure}
\centerline{\epsffile{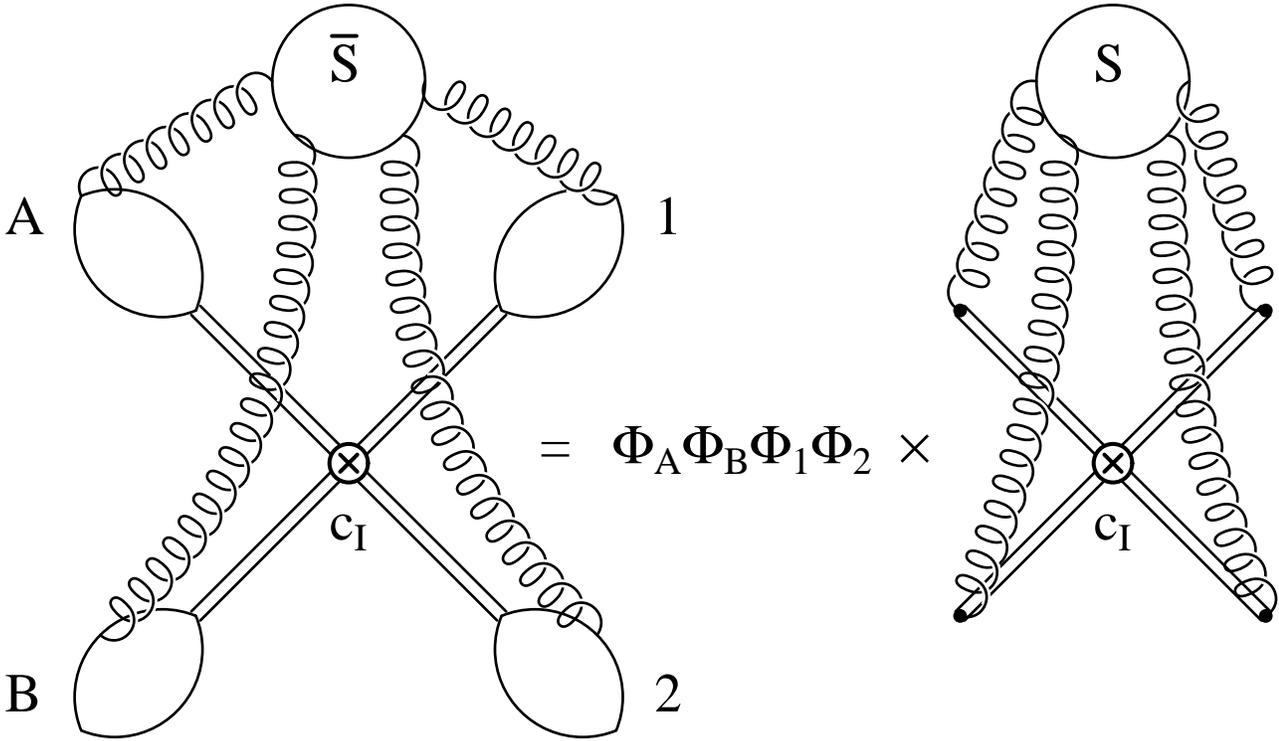}}
\caption{Identity representing factorization in
leading regions of corrections to the eikonal vertex $c_I$.
${\bar S}$ represents lines of zero momentum,  while
A, B, 1 and 2 label jet subdiagrams.}
\label{facteik}
\end{figure}

The relationship between the $W_I$ and the $S_I$ is shown in
Fig.\ \ref{facteik} in axial gauges.  
The left-hand side of the figure shows a general momentum configuration
of a ``leading region" \cite{CSSrv},
that contributes logarithmic behavior to (the vacuum expectation value of) $W_I^{(\a)}$;
lines have been grouped according to their momentum flow.  Such configurations
may be identified by standard Minkowski-space power counting \cite{CSSrv}.
As indicated in the diagram, all lines fall either into ``jet-like" subdiagrams,
labelled 1, 2, $A$, $B$, or into a ``soft" subdiagram, labelled $S$.
The jet subdiagrams consist of lines that are all on-shell with
light-like momenta parallel to the directions of the corresponding $\beta_i$.
The momenta of lines in $S$ vanish in all four components.  As indicated
in the figure, in each such ``leading" region, all collinear contributions
factor into the self energies, in axial gauges.  Demonstrations of this factorization,
based on Ward identities in QCD
may be found in Refs.\ \cite{CoSo81}, \cite{CSS1} and \cite{CSSrv} in the case of jets
of partons.  The eikonal lines in Fig.\ \ref{facteik} obey the same Ward identities
as do quark lines, and do not change these arguments.   

Under renormalization, the set of operators $W_I^{(\a)}$,
and hence the $S_I^{(\a)}$, will mix
in general.  Because the eikonal lines do not include
recoil effects, they cannot annihilate, and there is no
mixing between vertices that link other sets of Wilson
lines (corresponding to other flavor flows $\a$).  The
corresponding renormalization group equation is now in
terms of a matrix of anomalous dimensions,
\beqa
\mu{d\over d\mu} W_I^{(\a)} &=& 
(\Gamma_W^{(\a)})_{JI}\left(\alpha_s,\beta_i,\beta_j\right)\; 
W_J^{(\a)}\nonumber\\
\mu{d\over d\mu} S_I^{(\a)} &=& 
(\Gamma_S^{(\a)})_{JI}
\left(\alpha_s,{\beta_i\cdot\beta_j\; |n^2|\over |\beta_i\cdot n| \, 
| \beta_j\cdot n |}\right)\; S_J^{(\a)},
\label{Gammawijdef}
\eeqa
where in the arguments of $\Gamma_S$, $i\ne j$.
The two anomalous dimension matrices are related by a
generalization of Eq.\ (\ref{GammaSDYdef}) for Drell-Yan,
\beq
\Gamma_{S,IJ}^{(\a)}(\alpha_s,\beta_i,n) =
\Gamma_{W,IJ}^{(\a)}\left(\alpha_s,\beta_i,\beta_j\right)
-\delta_{IJ}\; \sum_{i=1}^4 
\Gamma_\Phi^{(f_i)}\left(\alpha_s,{|\beta_i\cdot n| \over \sqrt{\beta_i^2 |n^2|}}\right)\, .
\label{GammaSIJdef}
\eeq
The anomalous dimensions for $W_I^{(\a)}$ 
include collinear divergences in the limit
of vanishing $\beta_i^2$.  
These divergences, however, are
universal, and may be absorbed into the renormalization of
the $\Phi$'s.  This is possible precisely because any collinear
divergence may be absorbed into a self-interaction of one
of the Wilson lines, as argued above.  Since the renormalization of
self-energy graphs does not mix colors at the vertex,
collinear divergences in $\Gamma_W^{(\a)}$, Eq.\ (\ref{Gammawijdef}),
can only appear times the identity in the 
space of color tensors.
Therefore the soft anomalous dimension matrix $\Gamma_{S,IJ}^{(\a)}$ summarizes the mixing of operators
under renormalization, but is free from collinear divergences,
as long as $\beta_i\cdot \beta_j\ne 0$ for $i\ne j$.  
The soft anomalous dimension $\Gamma_S^{(\a)}$ will be, as in the case of  $\Gamma_S^{\rm (DY)}$ 
above,  free of universal collinear divergences, not only at one loop,
but to all orders in perturbation theory.

Note that the choice of finite renormalization for
the $\Phi$'s, and of overall gauge, encountered above,
may be interpreted as
a  renormalization scheme for defining the composite vertex
and its soft anomalous dimension.   In the next section, we  shall
exhibit a class of infrared safe soft functions
that occur in the threshold resummation of dijet
cross sections \cite{KOS1}, and whose evolution is controlled
by the tensors $\Gamma_{S,IJ}^{(\a)}$.

\section{Soft Functions in Dijet Production}

In this section we recall the role that
eikonal operators like those discussed above play
in the resummation of threshold corrections in
dijet production,
\beq
h_A(p_A)+h_B(p_B){\longrightarrow}J_1(p_1)+J_2(p_2)+X(k)\, ,
\label{process}
\eeq
at fixed dijet invariant mass,
\beq
M^2_{JJ}=(p_1+p_2)^2\, ,
\label{jetmass}
\eeq
and at fixed rapidity difference,
\beq
\Delta y = {1\over 2}\ln \left({p_1^+p_2^-\over p_1^-p_2^+}\right)=\ln\left({u\over t}\right)\, ,
\label{rapdiff}
\eeq
with $t$ and $u$ the usual Mandelstam variables, defined as
\beqa
t&=&\left(p_A-p_1 \right)^2 \nonumber \\
u&=&\left(p_A-p_2 \right)^2 \,.
\eeqa
As discussed in Ref.\ \cite{KOS1}, many other choices of cross section
involve essentially the same physics.  Again to be specific,
we assume that the two final-state jets in (\ref{process}) are identified by cones of opening
angles $\delta_{i}$.  

We shall be concerned with the contribution to this
cross section from an underlying partonic scattering  like
Eq.\ (\ref{sctgprocess}). In particular,
we shall review some relevant features of
the resummation, in this cross section, of ``threshold" corrections
to the hard-scattering function.  These are corrections of
the general form $[\ln^n(1-z) /(1-z)]_+$, with $z=M_{JJ}^2/\hat{s}$,
where $\hat{s}$ is the invariant mass squared of the partons
that initiate the process.  Referring to Eq.\ (\ref{sctgprocess}),
we take $\ell_A=x_Ap_A$, $\ell_B=x_Bp_B$, with $p_A$ and $p_B$ the
momenta of the incoming hadrons and $x_{A,B}$ momentum
fractions, so that $\hat{s}=2x_Ax_Bp_A\cdot p_B$.
We represent the coupling of soft gluons to the hard partons
of the flavors
$f_A$, $f_B$, $f_1$ and $f_2$ 
by Wilson lines, as in Eq.\ (\ref{eq:wilson}), tied together at 
the vertices of Eq.\ (\ref{eq:wivertex}).  
In terms of the operators $W^{(\a)}_I$ we define a dimensionless ``eikonal cross section", 
describing the emission
of gluons by hard partons in eikonal approximation, 
\beqa
\sigma_{LI}^{(\a,{\rm eik})}\left(\frac{wM_{JJ}}{\mu},\Delta y,\alpha_s(\mu^2),\delta_i,\epsilon\right)
&=&
\sum_{\xi}\, \delta\left(w-w^{({\rm eik})}(\xi,\delta_i)\right)
\nonumber\\
&\ & \hspace{-5mm}\times 
\langle0|{\bar T}\left[ \left(W^{(\a)}_L(0)\right){}^{\dagger}_{\{b_i\}}\right]|\xi{\rangle}
{\langle}\xi|T\left[W_I^{(\a)}(0)_{\{b_i\}}\right]|0 \rangle \, ,\nonumber \\
\label{eq:eikcs}
\eeqa
where $\xi$ designates a set of intermediate states.
The eikonal cross section depends on the rapidity difference $\Delta y$ 
(or equivalently, the partonic center-of-mass scattering angle) through
the relative directions of the Wilson lines 
coupled at each operator $W^{(\a)}$.
We choose the dimensionless weight $w$ in Eq.\ (\ref{eq:eikcs})
to reproduce the phase space
available to soft gluons near partonic threshold ($z=1$) \cite{KOS1}.
For the cross section in question, this requires that we fix the total
energy emitted by the eikonal lines into
the final state, but  outside the 
cones $\delta_i$.   Denoting this energy as $k^0_\xi$, we have
\beq
w^{(\rm eik)}\left(\xi,\delta_i \right)={2k_\xi^0\over M_{JJ}}\, .
\eeq
To handle the collinear divergences of 
$\sigma^{(\a,{\rm eik})}$ due to initial state interactions,
and to avoid infinite energy emitted into the cones, we 
may think of  the 
eikonals as having small ``mass", $\beta_i^2$.  We shall not
need this regulator in any of our explicit calculations below, however.

As in the previous section, we want to isolate the contributions of
noncollinear gluons, which control color mixing.   For this
purpose we factorize the eikonal cross section, Eq.\ (\ref{eq:eikcs}),
in the same manner as a hadronic cross section \cite{CSSrv,KOS1}.
Initial-state collinear divergences can be factorized
into eikonal ``parton distributions",
which we label $j_{\rm IN}$.  Analogously, final-state collinear
enhancements may be factorized into outgoing ``jet" functions, $j_{\rm OUT}$.

The result of this procedure is
to write the eikonal cross section as a convolution
of an infrared safe soft function $S^{(\a)}_{LI}$
with the functions $j_{\rm IN}$ and $j_{\rm OUT}$,
\beqa
\sigma_{LI}^{(\a,{\rm eik})}\left(\frac{wM_{JJ}}{\mu},\Delta y,\alpha_s(\mu^2),\delta_i,\epsilon\right)
&=& 
\nonumber\\
&\ & \hspace{-45mm} 
\int_0^1 dw'_Adw'_Bdw'_1dw'_2dw'_{S}\; \delta\left(w-w'_1-w'_2-w'_A-w'_B-w'_S\right)
\nonumber\\
&\ &\hspace{-40mm}
\times \prod_{c=A,B} {j}^{(f_c)}_{\rm{IN}}\left({w'_cM_{JJ}\over \mu},\alpha_s(\mu^2),\epsilon\right)\;
\, 
\prod_{d=1,2} {j}^{(f_d)}_{\rm{OUT}}
\left({w'_dM_{JJ}\over \mu},\alpha_s(\mu^2),\delta_d\right)
\nonumber\\
&\ & \hspace{-40mm} \times {S}^{(\a)}_{LI}\left({w'_{S}M_{JJ}\over \mu},\Delta y,\alpha_s(\mu^2)\right)\, .
\label{Stojjsprime}
\eeqa
Here $w'_i=2k'_i{}^0/M_{JJ}$, with $k'_i{}^0$ the energy of partons 
emitted outside the cones, and associated
with the function labelled by $i=1,2,A,B,S$.
We are using the incoherence of these ``effective theories'', as mentioned
in the introduction.

For completeness, we give definitions for
the incoming eikonal jets \cite{KS,KOS1}: 
\beqa
j^{(f_i)}_{\rm{IN}}\left({w'_iM_{JJ}\over \mu},\alpha_s(\mu^2),\epsilon\right) 
&=&
{M_{JJ}\over 2\pi}
\int_{-\infty}^\infty dy_0\ {\rm e}^{-iw'_iM_{JJ}y_0} \nonumber \\
&\ & \hspace{-15mm} \times 
\langle 0|\; {\rm Tr}\bigg\{\; {\bar T}[\Phi^{(f_i)}_{\beta_i}{}^\dagger(0,-\infty;y)]
T[\Phi^{(f_i)}_{\beta_i}(0,-\infty;0)]\; \bigg\}\; |0\rangle\, , \nonumber \\ 
\label{eq:eikjet}
\eeqa
with $i=A,B$ and $y^\nu=(y_0,\vec{0})$ a vector at the spatial origin. 
The argument $\epsilon=4-D$,
in $D$ dimensions, denotes the presence of collinear singularities in $j^{(f_i)}_{\rm{IN}}$.
Similarly, for the outgoing eikonal jets we have \cite{KOS1}, 
\beqa
j^{(f_i)}_{{\rm{OUT}}}\left({w'_iM_{JJ}\over \mu},\alpha_s(\mu^2),\delta_i\right) &=&
\sum_{\xi}\; \delta\left(w'_i-w_i^{({\rm eik})}(\xi,\delta_i)\right)
\nonumber\\
&\ & \hspace{-25mm} \times 
\langle 0|\; {\rm Tr}\bigg\{\; {\bar T}[\Phi^{(f_i)}_{\beta_i}{}^\dagger(\infty,0;0)]
|\xi \rangle \langle \xi|
 T[\Phi^{(f_i)}_{\beta_i}(\infty,0;0)]\; \bigg\}\; |0\rangle\, , \nonumber \\ 
\label{eq:eikoutjet}
\eeqa
with $i=1,2$. In this case, collinear singularities cancel due to the jet
definition in the sum over final states, and the infrared regulator $\epsilon$ 
is replaced by the cone size $\delta_i$.

Mellin transforms of the eikonal cross section with respect to the weight
conveniently isolate contributions from noncollinear soft gluons that
are singular at partonic threshold \cite{KOS1}.
If we take the Mellin transform of Eq.\ (\ref{Stojjsprime}),
\beq
\int_0^1 dw \, \left( 1-w \right)^{N-1} \, \sigma_{LI}^{(\a,{\rm eik})}
\left(\frac{wM_{JJ}}{\mu},\Delta y,\alpha_s(\mu^2),\delta_i,\epsilon\right) \, ,
\eeq
the convolution in the eikonal weights
becomes a product of moments, up to corrections that vanish as a power of  
the moment variable $N$ for large $N$. 
Hence the moments of the soft function can be extracted
from the eikonal cross section, as follows,
\beqa
{\tilde S}^{(\a)}_{LI}\left({M_{JJ}\over N\mu},\Delta y,\alpha_s(\mu^2)\right)
&=&
{{\tilde \sigma_{LI}}^{(\a,{\rm eik})}\left({M_{JJ}\over N\mu},\Delta y,\alpha_s(\mu^2),\delta_i,\epsilon\right)
\over
{\tilde j}^{(f_A)}_{\rm{IN}}\left({M_{JJ}\over N\mu},\alpha_s(\mu^2),\epsilon\right)\; 
{\tilde j}^{(f_B)}_{\rm{IN}}\left({M_{JJ}\over N\mu},\alpha_s(\mu^2),\epsilon\right)}\nonumber\\
&\ & \times
\frac{1}
{{\tilde{j}}^{(f_1)}_{{\rm{OUT}}}\left(\frac{M_{JJ}}{N\mu},\alpha_s(\mu^2),\delta_1\right) \: 
{\tilde{j}}^{(f_2)}_{{\rm{OUT}}}\left(\frac{M_{JJ}}{N\mu},\alpha_s(\mu^2),\delta_2\right)}\, .
\nonumber\\
\label{sigeikDY}
\eeqa
All the leading power singularities, 
$\left[\frac{\ln^m \left( 1-z \right)}{1-z}\right]_+$,
may be reconstructed from 
logarithms of the leading power (${\cal{O}}(N^0)$) $N$ dependence.

We recognize Eq.\ (\ref{sigeikDY}) as a generalization of 
Eq.\ (\ref{DYirredsoft}) to
the moments of the cross section. By removing the mutually-incoherent 
jet factors,
which in axial gauge correspond simply to cut parton self-energies, we
isolate in the function $S_{LI}^{(\a)}$ the interference of amplitudes
in which gluons are emitted by one eikonal line and absorbed by another.
In axial gauge, cut diagrams of this topology incorporate what we might
describe as ``interjet" coherent \cite{CoCoh} radiation in the cross section.

We have not yet discussed the renormalization of 
the soft function and eikonal jets. In the next section we will see that  
renormalization leads to the exponentiation of the $N$-dependence of the soft function,
and hence of the energy dependence of its transform.

\begin{figure}
\centerline{\epsffile{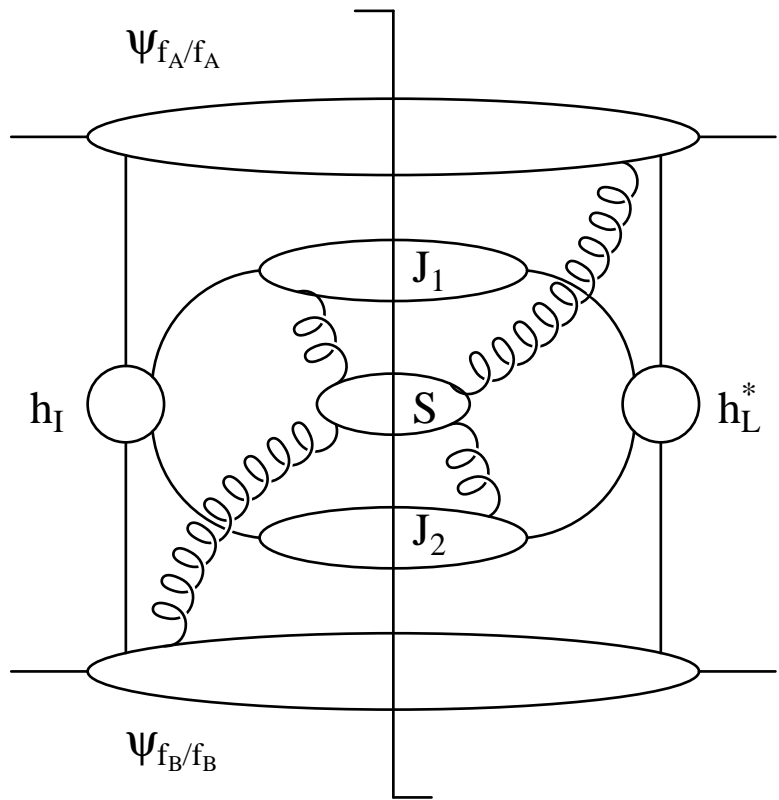}}
\caption{Reduced diagram representing the leading
region for the dijet cross section.}
\label{factocross}
\end{figure}

Now let us relate the soft function $\tilde{S}_{LI}$ to threshold
resummation for dijet cross sections.
To derive singular behavior at partonic threshold, we consider moments with 
respect to $\tau=M_{JJ}^2/S$ for
{\it perturbative}, partonic cross sections, in which the external
hadrons are replaced by partons in an infrared-regulated theory.  In Ref.\ \cite{KOS1}, we showed
that, up to next-to-leading logarithms, 
this cross section factorizes under such a Mellin transform, 
\beqa
\int_0^1 d\tau\; \tau^{N-1}\; S^2\;
\frac{d\sigma_{f_Af_B{\rightarrow}J_1J_2}(S,\delta_1,\delta_2)}
{dM^2_{JJ}d{\Delta}y} &=&         
\sum_{\a}\sum_{IL} 
H_{IL}^{(\a)}\left({M_{JJ}\over\mu},\Delta y,\alpha_s(\mu^2)\right)\nonumber\\
&\ &\ \hspace{-40mm}\times\; {\tilde\psi}_{f_A/f_A}\left ( N,{M_{JJ}\over \mu },\alpha_s(\mu^2),\epsilon \right )
\;
{\tilde\psi}_{f_B/f_B}\left (N,{M_{JJ}\over \mu },\alpha_s(\mu^2),\epsilon \right )  \nonumber\\
&\ &\ 
\hspace{-40mm}\times\; {\tilde S}_{LI}^{(\a)} \biggl ( {M_{JJ}\over \mu N},\Delta y,\alpha_s(\mu^2) \biggr ) 
\nonumber \\ 
&\ &\ 
\hspace{-40mm}\times\; {\tilde J}^{(f_1)}\left(N,{M_{JJ}\over \mu },\alpha_s(\mu^2),\delta_1\right)\; 
{\tilde J}^{(f_2)}\left(N,{M_{JJ}\over \mu },\alpha_s(\mu^2),\delta_2\right) \, .\nonumber \\
\label{eq:factimp}
\eeqa
The factorized cross section is illustrated in cut diagram notation in Fig.\ \ref{factocross}.
Here, $S_{LI}$ is exactly the matrix of soft eikonal functions constructed 
above.  $H_{IL}$ absorbs quanta, represented by $h_L^*$ and $h_I$ (see also Eq.\ (\ref{uvjetfact}) below), 
which are off-shell by the order of $M_{JJ}$.  
The remaining functions $\psi$ and $J$ include, respectively, virtual and final-state partons parallel to
the incoming hadrons (partons) and to the observed outgoing jets.  Explicit definitions
of the jet functions
are given in Ref.\ \cite{KOS1}.  Their detailed features are not of interest here, primarily
because they are color diagonal, and incoherent with the soft radiation included in $S_{LI}$.
Our interest is in the interplay of color structures between the soft tensor $S_{LI}$ and
the hard color tensor $H_{IL}$.

\section{Color Evolution in Jet Cross Sections}

In this section, we exhibit the evolution of color exchange in
hard scattering.  Although our notation is taken from the dijet cross section
of the previous section, it is quite general.  

\subsection{Evolution for Soft and Hard Functions}

The composite operators of the soft function ${{\tilde S}_{LI}}^{(\a)}$, as we saw before,
require
 renormalization.
The matrices $H$ and $S$ occur in a product in Eq.\ (\ref{eq:factimp})
and (consistent with the discussion of Sec.\ 2), must therefore renormalize
multipicatively, with separate renormalization factors
for the amplitude and the complex conjugate \cite{BottsSt,CLS},
\beqa
H^{(\a)}{}^{(0)}_{IL}&=& \prod_{i=A,B,1,2}Z_i^{-1}\; \left(Z_S^{(\a)}{}^{-1}\right)_{IC}H_{CD}^{(\a)}\; 
[(Z_S^{(\a)}{}^\dagger)^{-1}]_{DL}\cr
S^{(\a)}{}^{(0)}_{LI}&=&(Z_S^{(\a)}{}^\dagger)_{LB}S_{BA}^{(\a)}Z_{S,AI}^{(\a)}{}\, .
\label{eq:barereno}
\eeqa
Here  $Z_i$ is the wavefunction renormalization
constant of the {\em{i}}th incoming parton field (of flavor
$f_A$ \dots $f_2$) in Eq.\ (\ref{sctgprocess}),
involved in the hard scattering, and  $Z^{(\a)}_{S,CD}$ is a matrix of
 renormalization constants, describing
the renormalization of the soft function. 

In the following, we shall adopt the convention of using Roman indexes to refer to
the color structures
in an arbitrary basis. In jet cross sections
near partonic threshold, the hard-scattering function is 
a product of two purely virtual factors \cite{KOS1}
\beq
H_{IL}^{(\a)}\left({M_{JJ}\over\mu},\Delta y,\alpha_s(\mu^2)\right) 
=
{h_{L}^{(\a)}}^*\left({M_{JJ}\over\mu},\Delta y,\alpha_s(\mu^2)\right)\;
h_{I}^{(\a)}\left({M_{JJ}\over\mu},\Delta y,\alpha_s(\mu^2)\right)\, ,
\nonumber\\ 
\label{uvjetfact}
\eeq
one from the overall amplitude and one from the complex conjugate.
This simplification \cite{CoSo81} is due to the fact that near
threshold there is insufficient phase space for the emission
of hard partons, aside from those observed in the hard scattering
(and hence summarized in the jet functions $J^{(f)}$ in Eq.\ (\ref{eq:factimp})).

The $h_I$ in Eq.\ (\ref{uvjetfact}) are the coefficients of color tensors in the
expansion of the hard scattering amplitude $h^{\a)}_{d_A\dots d_2}$
treated as a matrix in color space,
\beq
h^{(\a)}_{d_A\dots d_2}\left({M_{JJ}\over\mu},\Delta y,\alpha_s(\mu^2)\right)
 =\sum_{K} h_{K}^{(\a)}\left({M_{JJ}\over\mu},\Delta y,\alpha_s(\mu^2)\right)\; 
\left(c_K^{(\a)}\right)_{d_A\dots d_2}\, ,
\label{hardsctgexpand}
\eeq
with the $c_K^{(\a)}$ the same color tensors introduced to
define the soft functions, through Eqs.\ (\ref{eq:wivertex}), (\ref{eq:eikcs}) and (\ref{sigeikDY}).
Here we may point out a strong similarity to the effective theory formalism,
in which the $c_I$'s play the role of operators in a ``low-energy"
theory, and the $h_I$'s are coefficient functions that summarize the
high-energy components that have been ``integrated out''.

From Eq.~(\ref{eq:barereno}), the soft function $S^{(\a)}_{LI}$ satisfies 
the renormalization group equation \cite{BottsSt,KS}
\beq 
\left(\mu\frac{\partial}{\partial\mu}+\beta(g)\frac{\partial}{{\partial}g}\right)S^{(\a)}_{LI}=
-(\Gamma_S^{(\a)}{}^\dagger)_{LB}S^{(\a)}_{BI}-S^{(\a)}_{LA}(\Gamma^{(\a)}_S)_{AI}\, ,
\label{eq:resoft}
\eeq
where we encounter the same {\em{soft anomalous dimension matrix}}, $\Gamma^{(\a)}_S$,
as in Eq.\ (\ref{Gammawijdef}), which is 
computed directly from the
UV divergences of the soft function. We will calculate this matrix for the basic
partonic processes $\a$ in a minimal 
subtraction renormalization scheme,
taking $\epsilon=\epsilon_{UV}=4-D$, with $D$ the number of space-time dimensions.
The one-loop anomalous dimensions, 
\beq
\Gamma^{(\a)}_S (g)=-\frac{g}{2} \frac {\partial}{{\partial}g}{\rm Res}_{\epsilon 
\rightarrow 0} Z^{(\a)}_S (g, \epsilon)\, ,
\label{eq:gendefano}
\eeq
are obtained from the residues of the UV poles contained in the 
matrices of renormalization constants of the soft function.

The moment dependence of the product of the hard and soft functions
may now be determined
directly from Eq.\ (\ref{eq:resoft}) \cite{KOS1},
\beqa
{\rm Tr}\Bigg\{ 
H^{(\a)}\left({M_{JJ}\over\mu},\Delta y,\alpha_s(\mu^2)\right)\;
{\tilde S}^{(\a)} \biggl ({M_{JJ}\over N\mu},\Delta y,\alpha_s(\mu^2) \biggr )\Bigg\}
&\ &\nonumber\\
&\ & \hspace{-70mm}
= {\rm Tr}\Bigg\{ 
H^{(\a)}\left({M_{JJ}\over\mu},\Delta y,\alpha_s(\mu^2)\right)\nonumber\\
&\ & \hspace{-70mm}\ \  \times 
\bar{P} \exp \left[\int_\mu^{M_{JJ}/N} {d\mu' \over \mu'}\; 
\Gamma_S^{(\a)}{}^\dagger\left(\alpha_s(\mu'^2)\right)\right]\nonumber\\
&\ & \hspace{-70mm}\ \  \times
{\tilde S}^{(\a)} \biggl (1,\Delta y,\alpha_s\left(M_{JJ}^2/N^2\right) \biggr )\nonumber\\
&\ & \hspace{-70mm}\ \  \times 
P \exp \left[\int_\mu^{M_{JJ}/N} {d\mu' \over \mu'}\; \Gamma_S^{(\a)}\left(\alpha_s(\mu'^2)\right)\right]
\Bigg\}\, ,
\label{softsoln}
\eeqa
where the symbols $P$ and $\bar{P}$ refer
to path-ordering in the same  and in the opposite sense as the integration variable $\mu'$.
$P$ orders $\Gamma_S^{(\a)}(\alpha_s(\mu^2))$ to the far right
and $\Gamma_S^{(\a)}(\alpha_s(M_{JJ}^2/N^2))$ to the far left of the products.
In Eq.\ (\ref{softsoln}) the trace is taken in the basis of 
color structures, as 
\beq
{\rm Tr}\Bigg\{ 
H^{(\a)}
{\tilde S}^{(\a)}\Bigg\} \equiv H^{(\a)}_{IL} \tilde{S}^{(\a)}_{LI}
={h^{(\a)}_L}^* \tilde{S}^{(\a)}_{LI} h^{(\a)}_I.
\label{eq:HS}
\eeq 
At lowest order, $\tilde{S}^{(\a)}_{LI}$ is just the square of eikonal vertices, 
with elements given by
color traces of $c_I$ and $c_L^*$.

\subsection{Diagonalization of the Color Basis} 
\label{chabasexp}

In lowest (linear) order in $\alpha_s$ for the anomalous
dimension matrices, the ordered exponentials of Eq.\ (\ref{softsoln})
may be reduced to sums of exponentials, by changing the color
basis to one in which $\Gamma_S^{(\a)}$ is diagonal.  We proceed as follows.

First, we treat 
the original basis of color structures as a set of kets, 
$\left\{ |{c^{(\a)}}_I \rangle \right\}$.  
Second, we compute the anomalous dimension matrix $\Gamma^{(\a)}_S$
at one loop in this basis and we then diagonalize it, finding its eigenvalues and eigenvectors. 
Let the diagonal basis of the eigenvectors
be denoted as the set $\left\{ |{e^{(\a)}}_{\kappa} \rangle \right\}$, where  
from now on we  shall use  Greek indices for 
vectors and matrices expressed in this basis.
The matrix defined as 
\beq
{\left( R^{(\a)} \right)}_{K \beta}^{-1} \equiv \langle  {c^{(\a)}}_K | 
{e^{(\a)}}_{\beta} \rangle \equiv \left( {e^{(\a)}}_{\beta} \right)_K,
\label{eq:Rmatr}
\eeq 
diagonalizes the anomalous dimension matrix according to the equation
\beq
\left[ R^{(\a)} \Gamma_{S}^{(\a)} \left( R^{(\a)} \right)^{-1} \right]_{\kappa \beta}=
{\lambda}_{\kappa}^{(\a)} \delta_{\kappa \beta}.
\label{eq:Rdiag}
\eeq
Third, we reexpress the hard amplitudes in the 
eigenvector basis. In the previous notation,
\beqa
\left( h^{(\a)}_{\gamma} \right)^* &=&  \left( h^{(\a)}_K \right)^* 
 {\left( R^{(\a)} \right)}_{K \gamma}^{\dagger}
 \nonumber \\
h^{(\a)}_{\beta}  &=&  {\left( R^{(\a)} \right)}_{ \beta L} h^{(\a)}_L\, .
\label{eq:newh} 
\eeqa
We reexpress Eq.\ (\ref{eq:HS}) in the new basis as
\beq
{\rm Tr}\Bigg\{ 
H^{(\a)}
{\tilde S}^{(\a)}\Bigg\} =\left( h^{(\a)}_{\gamma} \right)^*
\tilde{S}^{(\a)}_{\gamma \beta} 
h^{(\a)}_{\beta}=H^{(\a)}_{\beta \gamma} \tilde{S}^{(\a)}_{\gamma \beta}\,  ,
\label{eq:newHS} 
\eeq
where we define
\beq
\tilde{S}^{(\a)}_{\gamma \beta} \equiv 
\left[ \left(R^{(\a)}_{\gamma M} \right)^{-1} \right]^{\dagger}
\tilde{S}^{(\a)}_{MN}
{\left( R^{(\a)} \right)}_{N \beta}^{-1}\, .
\label{eq:newbasS}
\eeq
In the diagonal basis of the eigenvectors, Eq.\ (\ref{softsoln}) then becomes
\beqa
{\rm Tr}\Bigg\{H^{(\a)}\left({M_{JJ}\over\mu},\Delta y,\alpha_s(\mu^2)\right)\;
{\tilde S}^{(\a)} \biggl ({M_{JJ}\over N\mu},\Delta y,\alpha_s(\mu^2) \biggr )\Bigg\}
&\ &\nonumber\\
&\ & \hspace{-80mm}
= 
H^{(\a)}_{\beta \gamma}\left({M_{JJ}\over\mu},\Delta y,\alpha_s(\mu^2)\right)
{\tilde S}_{\gamma \beta} ^{(\a)} \biggl (1,\Delta y,\alpha_s\left(M_{JJ}^2/N^2\right) \biggr )\nonumber\\
&\ & \hspace{-80mm}\ \  \times 
\exp \Bigg\{ \int_\mu^{M_{JJ}/N} {d\mu' \over \mu'}\; \left[ \lambda^{(\a)*}_{\gamma}\left(\alpha_s(\mu'^2)\right)
+{\lambda}_{\beta}^{(\a)} \left(\alpha_s(\mu'^2)\right) \right] \Bigg\} \, , \nonumber\\
\label{softsoleigbas}
\eeqa
in which the leading $N$-dependence of the product is organized in a sum of exponentials.

In general, the eigenvalues in Eq.\ (\ref{softsoleigbas}) depend on the flavors and directions
of the colliding partons, both incoming and outgoing.
In the following two sections, for  the partonic processes $\a$,  we will give:
\begin{itemize}
\item  The basis of color structures $\left\{ |{c^{(\a)}}_I \rangle \right\}$. Generally, we will work with
       an arbitrary number of colors, $N_c$. 
\item  In this basis, the anomalous dimension matrix $\Gamma^{(\a)}_S$.
\item  The set of eigenvalues, $\left \{ {\lambda}_{\kappa}^{(\a)} \right \}$, and the
       corresponding set of eigenvectors, $\left\{ |{e^{(\a)}}_{\kappa} \rangle \right\}$,
       of $\Gamma^{(\a)}_S$.
       Each eigenvector, in the notation of Eq.\ (\ref{eq:Rmatr}), corresponds to a column of 
       ${\left( R^{(\a)} \right)}^{-1}$.
\end{itemize}
With these quantities, combined with explicit hard-scattering functions, it is
possible to write down resummed jet cross sections \cite{KOS1}.

The anomalous dimension matrices for the processes 
$q \bar{q}\rightarrow q \bar{q}$ and $q q \rightarrow q q$
have been known for some time \cite{BottsSt,SotiSt,GK,KK}.
In Sec.\ \ref{sec-qinipro} we will review these matrices at one loop, and specify their eigenvectors.
There we also give corresponding results for the process 
$q \bar{q}\rightarrow gg$ (related by time-reversal invariance
to $gg \rightarrow q \bar{q}$ as well \cite{KS,Thesis}) and for the process
$q g \rightarrow q g$ (related 
to $\bar{q} g \rightarrow  \bar{q} g$).
The quark-gluon processes involve a $3\times 3$ matrix, the quark-quark a $2\times 2$,
and all of these results are reasonably straightforward.
For gluon-gluon scattering
on the other hand, the anomalous dimension matrix is $8\times 8$ in $SU(3)$,
and requires a somewhat more extensive analysis, which we give in Sec.\ 6.

\section{Anomalous Dimensions for Quark-initiated Scattering}
\label{sec-qinipro}

\subsection{Formalism}
\label{sec-formal}

Our Wilson lines represent partonic processes, whose  momenta and colors 
are labeled as in Eq.\ (\ref{sctgprocess}),
$f_A\left( l_A, \e_A \right)+f_B\left( l_B, \e_B \right)$ $\rightarrow$ 
$f_1\left( p_1, \e_1 \right)+f_2\left( p_2, \e_2 \right)$.
In terms of hadronic momenta, we take  $l_i=x_i p_i$, $i=A,B$,
with $x_i$ the partonic momentum fraction. 
Referring to the dijet process Eq.\ (\ref{process}) at partonic threshold,
the dimensionless, lightlike velocity vectors $v_i^{\mu}$, $v_i^2=0$,
which define the directions of Wilson lines 
at the vertices $W^{(\a)}$, are given by
\beqa
l_i^{\mu}&=&M_{JJ} v_i^{\mu} \hspace{15mm} i=A,B \nonumber \\
p_i^{\mu}&=&M_{JJ} v_i^{\mu} \hspace{15mm} i=1,2 \, .
\eeqa
We also recall the definitions of partonic Mandelstam invariants, 
\beqa
\hat{s}&=&\left( l_A+l_B \right)^2 \nonumber \\
\hat{t}&=&\left( l_A-p_1 \right)^2 \nonumber \\
\hat{u}&=&\left( l_A-p_2 \right)^2 \, .
\label{Mandlst}
\eeqa
The anomalous dimension matrices for all the processes 
below depend on logarithms of ratios of these invariants, and we 
introduce for them the following notation:
\beqa
\T&\equiv& \ln(\frac{-\hat t}{\hat s})+i\pi\nonumber \\
\U&\equiv& \ln(\frac{-\hat u}{\hat s})+i\pi.
\label{eq:new2form}
\eeqa

As mentioned above, although
the full cross section is gauge independent, the functions
into which it is factorized depend, in general, on the choice of $n^{\mu}$,
the axial gauge-fixing vector.
From the factorized cross section, Eq.\ (\ref{eq:factimp}), 
it is clear that gauge dependence in the product of $H^{(\a)}$ and $S^{(\a)}$ must cancel the
gauge dependence of the incoming and outgoing jets, $\psi$ and $J^{(f_i)}$.
Because the jets are incoherent relative to the hard and soft functions, 
the gauge dependence
of the anomalous dimension matrices $\Gamma_S^{(\a)}$ must
be proportional to the identity matrix.
To summarize this gauge dependence we introduce, for each parton in process $\a$, the function
\beq
\gi=C_{f_i}\; {\alpha_s\over\pi} \left[-\frac{1}{2}
\ln(\nu_i)-\frac{1}{2}\ln2+\frac{1}{2}-\frac{1}{2}i\pi\right]\, ,
\label{eq:ovallgauge}
\eeq
with $C_{f_i}=C_F\ (C_A)$ for a quark (gluon), and with
(see also Appendix \ref{app-integ})
\beq
\nu_i\equiv\frac{(v_i \cdot n)^2}{|n|^2}\, .
\label{eq:gdepnu}
\eeq
In these terms, we will write our
anomalous dimension matrix as
\beq
(\Gamma^{(\a)}_S)_{KL}=(\Gamma^{(\a)}_{S'})_{KL}+\left(\sum_{i=A,B,1,2} \gi \right) \delta_{KL},
\label{gammagaug}
\eeq
and in the following, for specific processes, we will report only $\Gamma^{(\a)}_{S'}$,
to which $\Gamma^{(\a)}_S$ reduces under a proper choice of gauge.
Eq.\ (\ref{gammagaug}) will allow us to recover the full gauge-dependence, whenever necessary.
We emphasize here that the dependence of the anomalous dimension matrix elements on the external eikonal four-vectors
and on the axial gauge-fixing vector comes entirely from the 
evaluation of the graphs shown in Fig.\ \ref{fig_eiko}, 
no matter which partonic process we consider. The explicit computation of the graphs shown 
requires the eikonal Feynman rules, given in Appendix \ref{app-feyn},
and a few one-loop integrations, 
summarized in Appendix \ref{app-integ}.

One last remark has to be made about the choice of physical channel $s$, $t$, or $u$, for the
definition of the basis of color structures. All choices are allowed, and related to each other
by simple crossing transformations.
In the following, apart from the processes $q \bar{q}\rightarrow gg$ and 
$gg \rightarrow q \bar{q}$, which are better described in terms of $s$-channel color structures,
we will always use $t$-channel bases, which seems to be the
natural choice when analyzing forward scattering \cite{SotiSt}.
In the following four subsections we present the results for the anomalous dimension matrices 
$\Gamma^{(\a)}_{S'}$ for partonic processes involving quarks.

\subsection{Soft anomalous dimension for $q \bar{q}\rightarrow q \bar{q}$}

We treat the process
\beq
q\left( l_A, \e_A \right)+\bar{q}\left( l_B, \e_B \right)\longrightarrow 
q\left( p_1, \e_1 \right)+\bar{q}\left( p_2, \e_2 \right) \, ,
\eeq
in the $t$-channel singlet-octet color basis
\beqa
c_1&=&\delta_{\e_A,\e_1}\delta_{\e_B,\e_2}\nonumber\\
c_2&=&-\frac{1}{2N_c}\delta_{\e_A,\e_1}\delta_{\e_B,\e_2}+\frac{1}{2}\delta_{\e_A,\e_B}\delta_{\e_1,\e_2}.
\label{eq:basqqbar}
\eeqa 
The procedure we follow is the same as the one described in 
Refs.\ \cite{BottsSt} and \cite{KS,Thesis}.  
We evaluate one-loop corrections for each color vertex, 
as shown in Fig.\ \ref{fig_eiko}, 
determining in every  case the color decomposition
of its ultraviolet divergences.  The relevant scalar integrals
are reviewed in Appendix \ref{app-integ} below.   The UV divergences found with
the color vertex $c_I^{(\a)}$ in Fig.\ \ref{fig_eiko} in general generate 
counterterms for all the vertices $c_J^{(\a)}$ in the same set,
where $J$ may or may not equal $I$.  This is standard operator
mixing under  renormalization.  The elements of the anomalous
dimension matrix are then found from Eq.\ (\ref{eq:gendefano}).

In this manner, we find the matrix
\beq
\Gamma_{S'}=\frac{\alpha_s}{\pi}\left(
                \begin{array}{cc}
                 2{C_F}\T  &   -\frac{C_F}{N_c} \U  \\
                -2\U    &-\frac{1}{N_c}(\T-2\U)
                \end{array} \right)\, .
      \label{eq:gamfinqqbar}
\eeq
The dependence on the logarithmic ratio $\T$ is diagonal in a $t$-channel color basis. Referring
to the resummed $N$-dependence in Eq.\ (\ref{softsoleigbas}),
we see that in the forward region of the partonic scattering ($\T\rightarrow -\infty$) 
color singlet exchange is exponentially enhanced relative to color octet \cite{SotiSt}.
The eigenvalues of this anomalous dimension matrix are
\beqa
\lambda_1&=&\frac{\alpha_s}{\pi}\frac{1}{2N_c}\left[({N_c}^2-2)\T+2\U-N_c\sqrt{\Delta}\right]\nonumber\\
\lambda_2&=&\frac{\alpha_s}{\pi}\frac{1}{2N_c}\left[({N_c}^2-2)\T+2\U+N_c\sqrt{\Delta}\right]\, ,
\label{eq:eigvalqqbar}
\eeqa
where $\Delta$ is defined as
\beq
\Delta={N_c}^2{\T}^2-4\T\U+4{\U}^2.
\label{eq:Deltaqqbar}
\eeq
The corresponding (arbitrarily normalized) eigenvectors are
\beqa
e_1&=&
\left(
                \begin{array}{c}
\frac{-{N_c}^2\T+2\U+N_c\sqrt{\Delta}}{4N_c \U}\\ \vspace{2mm}
1
                \end{array} \right)\nonumber\\
e_2&=&
\left(
                  \begin{array}{c}

\frac{-{N_c}^2\T+2\U-N_c\sqrt{\Delta}}{4N_c \U}\\ \vspace{2mm}
1
                \end{array} \right)\, .\nonumber\\
\label{eq:eigvectqqbar}
\eeqa
From these eigenvectors we may reconstruct the matrix $R^{-1}$ of Eq.\ (\ref{eq:Rmatr}),
which gives the transformation from the  singlet-octet to the diagonal basis, 
for any scattering angle.
The remaining anomalous dimensions are found in much the same fashion.

\subsection{Soft anomalous dimension for $q q\rightarrow q q$}

In this subsection we analyze the process
\beq
q\left( l_A, \e_A \right)+q\left( l_B, \e_B \right)\longrightarrow 
q\left( p_1, \e_1 \right)+q\left( p_2, \e_2 \right) \, ,
\eeq
again in the $t$-channel singlet-octet color basis
\beqa
c_1&=&-\frac{1}{2N_c}\delta_{\e_A,\e_1}\delta_{\e_B,\e_2}+\frac{1}{2}\delta_{\e_A,\e_2}\delta_{\e_B,\e_1}\nonumber\\
c_2&=&\delta_{\e_A,\e_1}\delta_{\e_B,\e_2}.
\label{eq:basqqqq}
\eeqa 
We find  the anomalous dimension matrix \cite{SotiSt}
\beq
\Gamma_{S'}=\frac{\alpha_s}{\pi}\left(
                \begin{array}{cc}
                -\frac{1}{N_c}(\T+\U)+2C_F \U  &  2\U \\
                 \frac{C_F}{N_c} \U    & 2{C_F}\T 
                \end{array} \right).
      \label{eq:gamfinqqqq}
\eeq
Once more, the dependence on the logarithmic ratio $\T$ is diagonal in the $t$-channel color basis, and again
the color singlet dominates the octet in the forward region.
The eigenvalues of this anomalous dimension matrix are
\beqa
\lambda_1&=&\frac{\alpha_s}{\pi}\frac{1}{2N_c}\left[({N_c}^2-2)\left(\T+\U \right)-N_c\sqrt{\Delta'}\right]\nonumber\\
\lambda_2&=&\frac{\alpha_s}{\pi}\frac{1}{2N_c}\left[({N_c}^2-2)\left(\T+\U \right)+N_c\sqrt{\Delta'}\right],
\label{eq:eigvalqqqq}
\eeqa
where $\Delta'$ is defined as
\beq
\Delta'={N_c}^2\left( \T-\U \right)^2+4\T\U,
\label{eq:Deltaqqqq}
\eeq
and the corresponding eigenvectors are
\beqa
e_1&=&
\left(
                \begin{array}{c}

\frac{-{N_c}^3 \left(\T-\U \right)-2N_c\U-{N_c}^2\sqrt{\Delta'}}{\left({N_c}^2-1 \right) \, \U}\\ \vspace{2mm}
1
                \end{array} \right)\nonumber\\
e_2&=&
\left(
                  \begin{array}{c}

\frac{-{N_c}^3 \left(\T-\U \right)-2N_c\U+{N_c}^2\sqrt{\Delta'}}{\left({N_c}^2-1 \right) \, \U}\\ \vspace{2mm}
1
                \end{array} \right).\nonumber\\
\label{eq:eigvectqqqq}
\eeqa

\subsection{Soft anomalous dimension for $q \bar{q}\rightarrow g g$ and
$g g \rightarrow q \bar{q}$}

Next, consider the process
\beq
q\left( l_A, \e_A \right)+\bar{q}\left( l_B, \e_B \right)\longrightarrow 
g\left( p_1, \e_1 \right)+g\left( p_2, \e_2 \right) \, ,
\eeq
in the $s$-channel color basis
\beqa
c_1&=&\delta_{\e_A,\e_B}\delta_{\e_1,\e_2}\nonumber\\
c_2&=&d^{\e_1 \e_2 c}{\left( T_F^c \right)}_{\e_B \e_A}\nonumber\\
c_3&=&if^{\e_1 \e_2 c}{\left( T_F^c \right)}_{\e_B \e_A},
\label{eq:basqqbargluglu}
\eeqa 
where the $T_F^c$'s are the generators of $SU(N_c)$ in the fundamental representation, while $f^{abc}$ and
$d^{abc}$ are the totally antisymmetric and symmetric $SU(N_c)$ invariant tensors respectively.
In this basis, we find the anomalous dimension matrix
\beq
\Gamma_{S'}=\frac{\alpha_s}{\pi}\left(
                \begin{array}{ccc}
                 0  &   0  & \U-\T  \\ \vspace{2mm}
                 0  &   \frac{C_A}{2}\left(\T+\U \right)    & \frac{C_A}{2}\left(\U-\T \right) \\ \vspace{2mm}
                 2\left(\U-\T \right)  & \frac{N_c^2-4}{2N_c}\left(\U-\T \right)  & \frac{C_A}{2}\left(\T+\U \right)
                \end{array} \right).
      \label{eq:gamfinqqbargluglu}
\eeq
The same anomalous dimension describes also the time-reversed process 
\cite{KS,Thesis}
\beq
g\left(p_1, \e_1 \right)+g\left( p_2, \e_2 \right) \longrightarrow 
\bar{q}\left( l_A, \e_A \right)+q\left( l_B, \e_B \right) .
\eeq

The eigenvalues of $\Gamma_{S'}$ are the solutions of the cubic equation
\beqa
\lambda^3-\frac{\alpha_s}{\pi}C_A \left( \T+\U \right) \lambda^2
+\left(\frac{\alpha_s}{\pi}\right)^2 \left[
\left( \frac{C_A}{2}\left( \T+\U \right) \right)^2
-\frac{N_c^2+4}{4} \left( \U-\T \right)^2\right]\lambda
\nonumber \\ 
+\left(\frac{\alpha_s}{\pi}\right)^3 
C_A \left( \T+\U \right) \left( \U-\T \right)^2
=0 \, ,
\eeqa
given by
\beqa
\lambda_1&=&\frac{\alpha_s}{\pi}\frac{1}{3}
\left(\X^{1/3}-\Y+C_A \left( \T+\U \right)\right) ,
\nonumber \\
\lambda_{2,3}&=&\frac{\alpha_s}{\pi}\frac{1}{3}\left[-\frac{1}{2}(\X^{1/3}-\Y)
+C_A \left( \T+\U \right) \pm \frac{1}{2} i \sqrt{3} (\X^{1/3}+\Y)\right] ,
\label{eigvalqqbargg}
\eeqa            
where
\beqa           
\X&=&-\left(\frac{C_A}{2}\left( \T+\U \right)\right)^3
+\frac{9}{8} C_A \left( \T+\U \right) \left(  \U-\T \right)^2 (N_c^2-8)
\nonumber \\ &&  
\hspace{-10mm}+\frac{3\sqrt{3}}{2} \left[
-\left( \U-\T \right)^6 \frac{(N_c^2+4)^3}{16}
-\left( \U-\T \right)^2 \left(\frac{C_A}{2}\left( \T+\U \right)\right)^4 (N_c^2-4)\right.
\nonumber \\ &&
\left.+\frac{1}{2} \left( \U-\T \right)^4 
\left(\frac{C_A}{2}\left( \T+\U \right)\right)^2 \left((N_c^2-8)^2-12(N_c^2-2)\right)
\right]^{1/2} \, ,
\label{Xqqbargg}
\eeqa         
and
\beq           
\Y=-\left[\left(\frac{C_A}{2}\left( \T+\U \right)\right)^2
+\frac{3}{4}\left( \U-\T \right)^2(N_c^2+4)\right]\X^{-1/3} .
\label{Yqqbargg}
\eeq         
For each eigenvalue $\lambda_i$
the corresponding eigenvector is given as
\beq            
e_i=\left(\begin{array}{c}
\frac{\U-\T}{\lambda_i} \vspace{3mm} \\ 
\frac{N_c \left( \U-\T \right)}{ 2\lambda_i-C_A\left(\T+\U \right)}\\ \vspace{2mm} 
1
\end{array}\right), \; i=1,2,3 .
\label{eigvecqqbargg}
\eeq          

\subsection{Soft anomalous dimension for $qg \rightarrow qg$ and
$\bar{q} g \rightarrow \bar{q} g$}

We now turn to the ``Compton'' process 
\beq
q\left( l_A, \e_A \right)+g\left( l_B, \e_B \right)\longrightarrow 
q\left( p_1, \e_1 \right)+g\left( p_2, \e_2 \right) \, ,
\eeq
in the $t$-channel color basis
\beqa
c_1&=&\delta_{\e_A,\e_1}\delta_{\e_B,\e_2}\nonumber\\
c_2&=&d^{\e_B \e_2 c}{\left( T_F^c \right)}_{\e_1 \e_A}\nonumber\\
c_3&=&if^{\e_B \e_2 c}{\left( T_F^c \right)}_{\e_1 \e_A}.
\label{eq:basqgqg}
\eeqa 
The result of our calculation is
\beq
\Gamma_{S'}=\frac{\alpha_s}{\pi}\left(
                \begin{array}{ccc}
                 \left( C_F+C_A \right) \T  &   0  & \U  \\ \vspace{2mm}
                 0  &   C_F \T+ \frac{C_A}{2} \U     & \frac{C_A}{2} \U \\ \vspace{2mm}
                 2\U  & \frac{N_c^2-4}{2N_c}\U  &  C_F \T+ \frac{C_A}{2}\U
                \end{array} \right),
      \label{eq:gamfinqgluqglu}
\eeq
which applies to the process
\beq
\bar{q}\left(p_1, \e_1 \right)+g\left( p_2, \e_2 \right)\longrightarrow
\bar{q}\left( l_A, \e_A \right)+g\left( l_B, \e_B \right) 
\eeq
as well.
As in Eq.\ (\ref{eq:gamfinqqbar}), the dependence on
the logarithmic ratio $\T$ is diagonal in 
the $t$-channel color basis, and the $t$-channel 
color singlet dominates in the forward region ($\T \rightarrow -\infty$).

The eigenvalues have the same structure as in Eq.\ (\ref{eigvalqqbargg}),
\beqa
\lambda_1&=&\frac{\alpha_s}{\pi}\frac{1}{3}
\left(\X'{}^{1/3}-\Y'+\left( 3C_F+C_A\right)\T+C_A\U \right) ,
\nonumber \\
\lambda_{2,3}&=&\frac{\alpha_s}{\pi}\frac{1}{3}\left[-\frac{1}{2}(\X'{}^{1/3}-\Y')
+\left(3C_F+C_A\right)\T+C_A\U \pm \frac{1}{2} i \sqrt{3} (\X'{}^{1/3}+\Y')\right] \, , \nonumber \\
\label{eigvalqgqg}
\eeqa        
with $\X'$ and $\Y'$ given by
\beqa           
\X'&=&\left(C_A \left( \T-\frac{\U}{2} \right)\right)^3
-\frac{9}{4} C_A \left( \T-\frac{\U}{2}\right) \U^2 (N_c^2-8)
\nonumber \\ &&  
+\frac{3\sqrt{3}}{2} \left[
-\U^6 \frac{(N_c^2+4)^3}{16}
-U^2 \left(C_A \left( \T-\frac{\U}{2} \right)\right)^4 (N_c^2-4)\right.
\nonumber \\ &&
\left.+\frac{1}{2} \U^4 
\left(C_A \left( \T-\frac{\U}{2} \right)\right)^2 \left((N_c^2-8)^2-12(N_c^2-2)\right)
\right]^{1/2}
\label{Xqgqg}
\eeqa         
and
\beq        
\Y'=-\left[\left(C_A \left( \T-\frac{\U}{2} \right)\right)^2
+\frac{3}{4}\U^2(N_c^2+4)\right]\X'{}^{-1/3} .
\label{Yqgqg}
\eeq         
The eigenvectors have a form similar to Eq.\ (\ref{eigvecqqbargg}),
\beq            
e_i=\left(\begin{array}{c}
\frac{\U}{\lambda_i-\left( C_F+C_A \right)\T} \vspace{3mm} \\  
\frac{N_c \U }{ 2\lambda_i-2C_F \T-C_A \U }\\ \vspace{2mm} 
1
\end{array}\right), \; i=1,2,3.
\label{eigvecqgqg}
\eeq          

\section{Soft anomalous dimensions for $gg \rightarrow gg$}

In this section, we present the anomalous dimension matrix for 
the eikonal version of gluon-gluon scattering.
Here the  proper choice of a set of
color vertices is  somewhat less obvious.  We would like
to compute, as easily as possible, the color structure of
one-loop corrections.
This was not a problem for $2\rightarrow 2$ partonic processes
involving quarks, because the couplings of the fundamental
representation are rather simple.

In subsection \ref{subsec-chobas} below, we introduce a basis of nine color tensors \cite{Macfar,Dixon},
which, although overcomplete, is convenient for computation of the one-loop
$\Gamma_{S'}$ for the gluon-gluon case.  
We give a few details of
how to color-decompose the one-loop graphs into this set, and present 
a first, ``raw'' version of the anomalous dimension matrix $\Gamma_{S'}$.
In subsection \ref{subsec-lintrasf}, we modify the original basis,
to one in which $\Gamma_{S'}$ is in a block-diagonal form
in which one block is diagonalized. 
We also present eigenvalues and eigenvectors in
this basis.
Finally, in subsection \ref{subsec-projbas}, 
working with $N_c=3$, we recall $t$-channel color projectors describing
the decomposition of the product of two color octets into irreducible representations \cite{Bartels}.
We use these projectors to reduce our set of color structures to a complete, eight-dimensional basis of 
color tensors. We exhibit the resulting one-loop anomalous dimension matrix,
along with its eigenvalues and eigenvectors.  As in all cases above,
logarithms of $t/s$ appear only in the diagonal of $\Gamma_{S'}$
in this basis.

\subsection{Initial basis and $\Gamma_{S'}$ at one loop}
\label{subsec-chobas}

\begin{figure}
\centerline{\epsffile{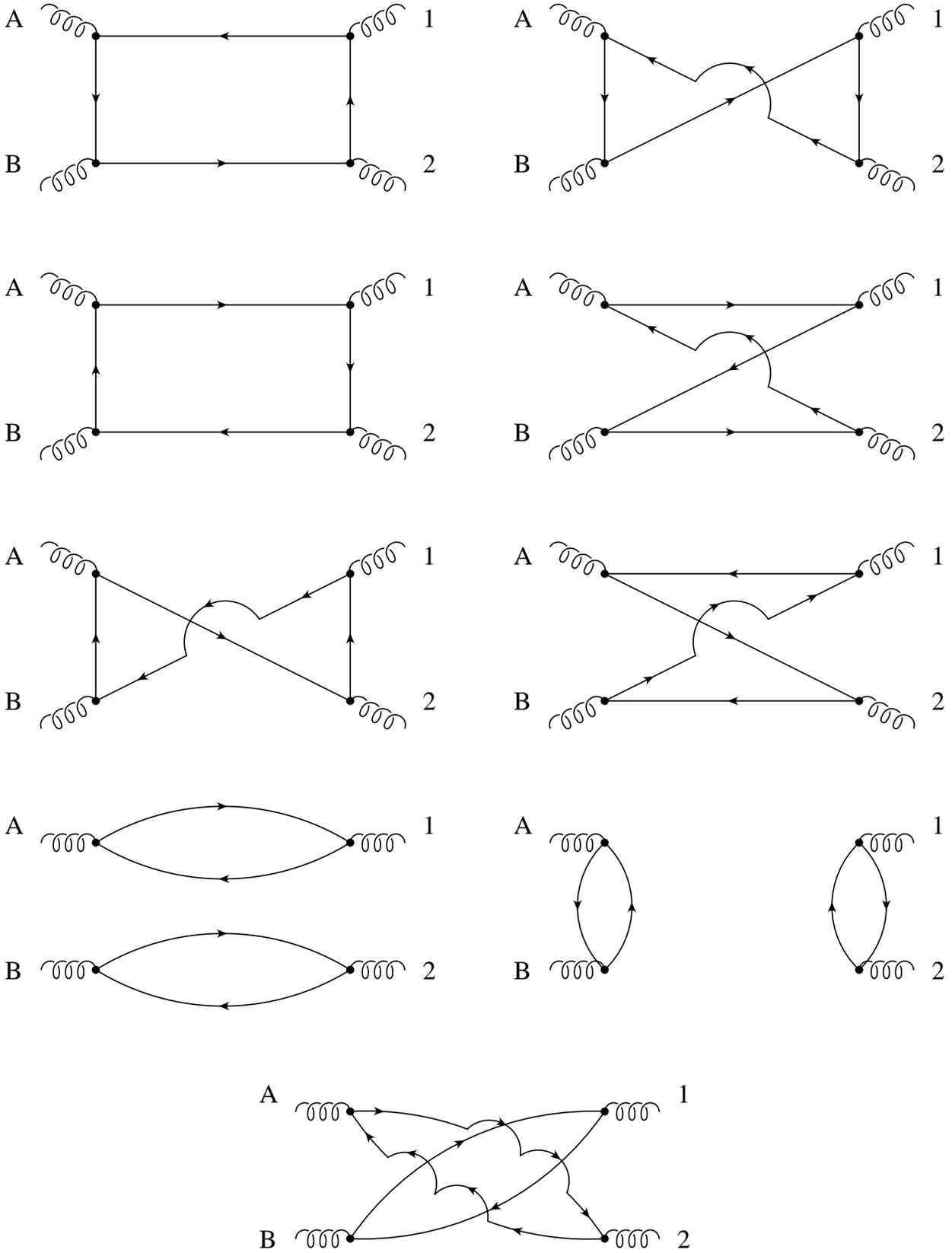}}
\caption{Graphical representation of the initial color basis for
the $gg \rightarrow gg$ process.  All lines have color
content only.}
\label{fig_ggcol}
\end{figure}

To begin with, we consider the set of color tensors, in terms of
which an arbitrary four-gluon diagram, with external color
indices $\e_A,\ \e_B,\ \e_1$ and $\e_2$ can be expanded \cite{Macfar,Dixon} 
\beqa
c_1&=&{\rm Tr}(T_{F}^{\e_A} T_{F}^{\e_B} T_{F}^{\e_2} T_{F}^{\e_1}) \, ,
\nonumber \\
c_2&=&{\rm Tr}(T_{F}^{\e_A} T_{F}^{\e_B} T_{F}^{\e_1} T_{F}^{\e_2}) \, ,
\nonumber \\
c_3&=&{\rm Tr}(T_{F}^{\e_A} T_{F}^{\e_1} T_{F}^{\e_2} T_{F}^{\e_B}) \, ,
\nonumber \\
c_4&=&{\rm Tr}(T_{F}^{\e_A} T_{F}^{\e_1} T_{F}^{\e_B} T_{F}^{\e_2}) \, ,
\nonumber \\
c_5&=&{\rm Tr}(T_{F}^{\e_A} T_{F}^{\e_2} T_{F}^{\e_1} T_{F}^{\e_B}) \, ,
\nonumber \\
c_6&=&{\rm Tr}(T_{F}^{\e_A} T_{F}^{\e_2} T_{F}^{\e_B} T_{F}^{\e_1}) \, ,
\nonumber \\
c_7&=&\frac{1}{4}\delta_{\e_A \e_1} \delta_{\e_B \e_2} \, ,
\nonumber \\
c_8&=&\frac{1}{4}\delta_{\e_A \e_B} \delta_{\e_1 \e_2} \, ,
\nonumber \\
c_9&=&\frac{1}{4}\delta_{\e_A \e_2} \delta_{\e_B \e_1} \, ,
\label{basgg1}
\eeqa
where the $T_{F}^{\e_i}$'s are the generators of $SU(N_c)$ in the  fundamental representation.
This set is illustrated graphically in Fig.\ \ref{fig_ggcol}, where all the lines and vertices have only color
content. The correspondence between the picture and Eq.\ (\ref{basgg1}) is straightforward, in terms of
the color dependence of the quark-gluon vertex in the QCD Lagrangian.

As in the case of scattering represented by Wilson lines in the
fundamental representation, there is no mixing of operators with 
different numbers of Wilson lines.  Therefore we only need to consider
color tensors with four indices in the adjoint representation.
Eq.\ (\ref{basgg1}) clearly includes the maximum number of inequivalent traces we can build out of
four generators $T_{F}^{\e_i}$, in addition to the three possible singlet combinations.
The set (\ref{basgg1}) mixes only with
itself under renormalization.
To see this, consider one-loop corrections to the vertices of Eq.\ (\ref{basgg1}). 
The color decomposition of these one-loop corrections is obtained by the graphical identities shown 
in Fig.\ \ref{fig_coldec}, whose analytic form is
\beqa
&&f^{\e_1 \e_2 \e_3}=-2i \left[ {\rm Tr} \left( T_{F}^{\e_1} T_{F}^{\e_2} T_{F}^{\e_3} \right)-
{\rm Tr} \left( T_{F}^{\e_1} T_{F}^{\e_3} T_{F}^{\e_2} \right) \right] \nonumber \\ 
&&\sum_{k}{\left[\left(T_{F}^k \right)_{\e_3 \e_1} \left(T_{F}^k \right)_{\e_4 \e_2}\right]}=
{1 \over 2} \delta_{\e_4 \e_1} \delta_{\e_3 \e_2} -{ 1 \over {2N_c}} \delta_{\e_3 \e_1} \delta_{\e_4 \e_2} \, ,
\label{breaktrace}
\eeqa
where $f^{\e_1 \e_2 \e_3}$ are the structure constants of the $SU(N_c)$ algebra. 
The first identity allows us to replace three-gluon couplings by
color traces in the fundamental representation, and the second to
``expand" the color content of
internal gluons and Wilson lines into pairs of color lines in the fundamental representation.
External gluon lines may be treated in the same way, by use of the identity
$\delta_{ab}=2{\rm Tr}\left[T_{F}^aT_{F}^b\right]$.
It is straightforward to extend this procedure iteratively to an arbitrary order,
using the corresponding identities for the four-point function. 
Essentially the same color decomposition can be used to show that 
the basis (\ref{basgg1}) is complete in the expansion of the hard-scattering
function for gluon-gluon scattering, as in Eq.\ (\ref{hardsctgexpand}) above.

As in Sec.\ 5, we follow the procedure of Refs.\ \cite{BottsSt} 
and \cite{KS,Thesis},
and compute the anomalous dimensions from the color decomposition 
into the set (\ref{basgg1}) of the
one-loop ultraviolet divergences of Fig.\ \ref{fig_eiko}, using Eq.\ (\ref{eq:gendefano}).
The choice of basis in Eq.\ (\ref{basgg1}) is particularly well-adapted 
to this procedure, when we employ the identities of Eq.\ (\ref{breaktrace}).
After straightforward, although rather tedious, calculations, we find the $9 \times 9$ anomalous dimension matrix, 
\beq
\Gamma_{S'}=\frac{\alpha_s C_A}{\pi} \left(
                \begin{array}{ccccccccc}
                  \T  &   0  & 0 & 0 & 0 & 0 & -\frac{\U}{N_c} &\frac{\T-\U}{N_c} & 0  \vspace{2mm} \\ 
                  0  &  \U & 0 & 0 & 0 & 0 & 0 & \frac{\U-\T}{N_c}& -\frac{\T}{N_c}    \vspace{2mm} \\
                  0  &  0  &  \T & 0 & 0 & 0 & -\frac{\U}{N_c} &\frac{\T-\U}{N_c} & 0  \vspace{2mm} \\ 
                  0 & 0 & 0 &  \left( \T+\U \right) & 0 & 0 & \frac{\U}{N_c} & 0 &\frac{\T}{N_c} \vspace{2mm} \\ 
                  0 & 0 & 0 & 0 &  \U &0 &0 & \frac{\U-\T}{N_c} &-\frac{\T}{N_c} \vspace{2mm} \\
                  0  & 0 & 0 & 0 & 0 &  \left( \T+\U \right) &\frac{\U}{N_c} &0 &\frac{\T}{N_c} \vspace{2mm} \\ 
                 \frac{ \T-\U}{N_c}  &0 & \frac{\T-\U}{N_c} &\frac{\T}{N_c} &0 &\frac{\T}{N_c} &2\T &0 &0 \vspace{2mm} \\ 
                  -\frac{\U}{N_c} &-\frac{\T}{N_c} & -\frac{\U}{N_c} & 0 
&-\frac{\T}{N_c} &0 &0 &0 &0  \vspace{2mm} \\ 
                  0 &\frac{ \U-\T}{N_c} & 0 &\frac{\U}{N_c} &\frac{\U-\T}{N_c} & \frac{\U}{N_c} & 0 & 0 &2  \U
                  \end{array} \right).
      \label{eq:gamfingggg}
\eeq
Interestingly, this anomalous dimension matrix is diagonal in the large $N_c$ limit.  

\subsection{Block-diagonal form of the anomalous dimension matrix:
eigenvalues and eigenvectors }
\label{subsec-lintrasf}

Although it has many zeros, the anomalous dimension matrix of Eq.\ (\ref{eq:gamfingggg})
is not transparently easy to diagonalize for arbitrary $N_c$.  As a first step in its simplification,
we will make a change of color basis that transforms  (\ref{eq:gamfingggg}) into a
block-diagonal form. 
In particular, looking at the first three rows  of
the matrix, we notice that, if we are able to eliminate the non-zero off-diagonal entries,
we immediately determine three of the eigenvalues.

The way to accomplish our purpose is 
to employ  symmetric and antisymmetric linear transformations from the set (\ref{basgg1}) to a new set
$\left \{ c_I' \right \}$, 
\beqa
c_1'&=&{\rm Tr}(T_{F}^{\e_A} T_{F}^{\e_B} T_{F}^{\e_2} T_{F}^{\e_1})-
{\rm Tr}(T_{F}^{\e_A} T_{F}^{\e_1} T_{F}^{\e_2} T_{F}^{\e_B})=c_1-c_3 \, ,
\nonumber \\
c_2'&=&{\rm Tr}(T_{F}^{\e_A} T_{F}^{\e_B} T_{F}^{\e_1} T_{F}^{\e_2})-
{\rm Tr}(T_{F}^{\e_A} T_{F}^{\e_2} T_{F}^{\e_1} T_{F}^{\e_B})=c_2-c_5 \, ,
\nonumber \\
c_3'&=&{\rm Tr}(T_{F}^{\e_A} T_{F}^{\e_1} T_{F}^{\e_B} T_{F}^{\e_2})-
{\rm Tr}(T_{F}^{\e_A} T_{F}^{\e_2} T_{F}^{\e_B} T_{F}^{\e_1})=c_4-c_6 \, ,
\nonumber \\
c_4'&=&{\rm Tr}(T_{F}^{\e_A} T_{F}^{\e_B} T_{F}^{\e_2} T_{F}^{\e_1})+
{\rm Tr}(T_{F}^{\e_A} T_{F}^{\e_1} T_{F}^{\e_2} T_{F}^{\e_B})=c_1+c_3 \, ,
\nonumber \\
c_5'&=&{\rm Tr}(T_{F}^{\e_A} T_{F}^{\e_B} T_{F}^{\e_1} T_{F}^{\e_2})+
{\rm Tr}(T_{F}^{\e_A} T_{F}^{\e_2} T_{F}^{\e_1} T_{F}^{\e_B})=c_2+c_5 \, ,
\nonumber \\
c_6'&=&{\rm Tr}(T_{F}^{\e_A} T_{F}^{\e_1} T_{F}^{\e_B} T_{F}^{\e_2})+
{\rm Tr}(T_{F}^{\e_A} T_{F}^{\e_2} T_{F}^{\e_B} T_{F}^{\e_1})=c_4+c_6 \, ,
\nonumber \\
c_7'&=&\frac{1}{4}\delta_{\e_A \e_1} \delta_{\e_B \e_2}=c_7 \, ,
\nonumber \\
c_8'&=&\frac{1}{4}\delta_{\e_A \e_B} \delta_{\e_1 \e_2}=c_8 \, ,
\nonumber \\
c_9'&=&\frac{1}{4}\delta_{\e_A \e_2} \delta_{\e_B \e_1}=c_9 \, .
\label{basgg2}
\eeqa
The expression of $\Gamma_{S'}$ in this new set of color structures has the form
\beq
\Gamma_{S'}=\left(
                \begin{array}{cc}
                 \blocA & 0_{3 \times 6} \\
                   0_{6 \times 3}    & \blocB
                  \end{array} \right),
      \label{eq:gamfinblo}
\eeq
where the block $\blocA$ is diagonal,
\beq
\blocA=\frac{\alpha_s}{\pi} \left(
                \begin{array}{ccc}
                  N_c\T  &   0  & 0  \\ 
                  0  &  N_c\U & 0    \\
                  0  &  0  &  N_c\left(\T+\U \right) 
                   \end{array} \right),
\label{eq:subbloc1}
\eeq
while the block $\blocB$ is given by
\beq
\blocB=\frac{\alpha _s}{\pi}  
\left(\begin{array}{cccccc}
N_c\T & 0 & 0 & -\U & \T-\U & 0 \\ 
0 & N_c\U & 0 & 0 & \U-\T & -\T \\ 
0 & 0 & N_c\left( \U+\T \right) & \U & 0 & \T \\ 
2\left( \T-\U \right) & 0 & 2\T & 2N_c\T & 0 & 0 \\ 
-2\U & -2\T & 0 & 0 & 0 & 0 \\ 
0 & -2\left( \T-\U \right) & 2\U & 0 & 0 & 2N_c\U 
\end{array}
\right). 
\label{eq:subbloc2}
\eeq
In both the previous matrices we have used explicitly $C_A=N_c$ (compare with
(\ref{eq:gamfingggg})).

The first three eigenvalues, $\lambda_1$, $\lambda_2$ and $\lambda_3$, can be read 
directly from Eq.\ (\ref{eq:subbloc1}). To determine the others, we just need
the diagonalization of a $6 \times 6$ matrix.
We have computed its eigenvalues and obtained
\beqa
\lambda _4&=&\frac{\alpha _s}{\pi} N_c\T=\lambda_1 ,
\nonumber \\ 
\lambda _5&=&\frac{\alpha _s}{\pi} N_c\U=\lambda_2 ,
\nonumber \\ 
\lambda _6&=&\frac{\alpha _s}{\pi} N_c\left( \T+\U \right)=\lambda_3 , 
\nonumber \\ 
\lambda _7&=&\frac{\alpha _s}{\pi} \left[X^{1/3}-Y+\frac{2}{3}N_c \left( \T+\U \right)\right] ,
\nonumber \\ 
\lambda _{8,9}&=&\frac{\alpha _s}{\pi} 
\left[-\frac{1}{2}(X^{1/3}-Y)+\frac{2}{3}N_c \left( \T+\U \right)
\pm\frac{1}{2}i\sqrt{3}\left( X^{1/3}+Y \right)\right] ,
\eeqa
where, as before, to reduce clutter in the notation, we introduce auxiliary quantities $\X$ and $\Y$,
defined for this particular case as
\beqa
\X&=&\frac{4}{27}N_c\left(N_c^2-9\right)\left[2\left( \T^3+\U^3\right)-3\left( \T^2\U+\T\U^2 \right)\right]
\nonumber \\ &&
+\frac{4}{9}\left\{-12\left( N_c^2-1\right)^2\left[ \T^6+\U^6-3\left( \T^5\U+\T\U^5 \right) \right]\right.
\nonumber \\ &&
-3\left( N_c^6+6N_c^4+33N_c^2+24 \right) \left( \T^4\U^2+\T^2\U^4 \right)
\nonumber \\ &&
\left. +6\T^3\U^3\left( N_c^6-4N_c^4+53N_c^2+14 \right)\right\}^{1/2}
\eeqa
and
\beq
\Y=-\frac{4}{9}\left( N_c^2+3 \right) \left[\T^2+\U^2-\T\U \right] X^{-1/3} .
\eeq
The eigenvectors corresponding to the first three eigenvalues have the form
\beq
e_i=\left(\begin{array}{c}
     e_i^{(3)} \\ 
      0^{(6)}
\end{array}\right), \; i=1,2,3 \, ,
\label{eigvct99}
\eeq
where the superscripts refer to the dimension. Thus, the $e_i^{(3)}$'s 
are three-dimensional vectors, defined as
\beq
e_i^{(3)}=\left(\begin{array}{c}
     \delta_{i1} \\ 
     \delta_{i2} \\  
        \delta_{i3} 
\end{array}\right), \; i=1,2,3 \, ,
\label{eq:3parteigv}
\eeq
while $0^{(6)}$ is the six-dimensional null vector.

The eigenvectors corresponding to the other eigenvalues have the general form
\beq
e_i=\left(\begin{array}{c}
     0^{(3)} \\ 
     e_i^{(6)} 
\end{array}\right), \; i=4,\ldots,9 \, .
\label{eq:6parteeigv}
\eeq
In particular $e_4^{(6)}$,  $e_5^{(6)}$ and 
$e_6^{(6)}$ are given by 
\beq
e_4^{(6)}=\left(\begin{array}{c}
\frac{\left( N_c^2-2 \right)\T}{2\U} \vspace{2mm} \\ 
1\\
\frac{\T-\U}{\U} \vspace{2mm}  \\
N_c \frac{\U-\T}{\U} \vspace{2mm} \\
-N_c  \vspace{2mm} \\
0
\end{array}\right) , \;  \;  \;
e_5^{(6)}=\left(\begin{array}{c}
-\frac{1}{N_c} \vspace{2mm}\\ 
-\frac{\left( N_c^2-2 \right) \U}{2N_c\T} \vspace{2mm}   \\
\frac{\T-\U}{\T N_c} \vspace{2mm}\\
0 \vspace{2mm}\\
1  \vspace{2mm} \\
\frac{\U-\T}{\T}
\end{array}\right) , \;  \; \;
e_6^{(6)}=\left(\begin{array}{c}
-\frac{\T}{\U} \vspace{2mm} \\ 
1 \vspace{2mm} \\
\left(N_c^2-2 \right)\frac{\U-\T}{2\U} \vspace{2mm} \\
N_c \frac{\T}{\U} \vspace{2mm} \\
0  \vspace{2mm} \\
-N_c
\end{array}\right) ,
\label{456eigvc}
\eeq
while $e_7^{(6)}$,  $e_8^{(6)}$ and 
$e_9^{(6)}$
are given in terms of the 
coefficients of $\alpha_s/\pi$ in the 
corresponding
eigenvalues, 
\beq
\lambda_i'=\frac{\pi}{\alpha_s} \lambda_i \, ,
\label{redeival}
\eeq
by  
\beq
e_i^{(6)}=\left(\begin{array}{c}
x_1(\lambda_i') \\ 
x_2(\lambda_i')\\
x_3(\lambda_i') \\
x_4(\lambda_i') \\
1  \\
x_6(\lambda_i')
\end{array}\right), \; i=7,8,9.
\label{789eigvc}
\eeq
Here, once more to shorten our rather complicated formulas, 
we have parametrized the eigenvectors in terms of the quantities 
\beqa
x_1(\lambda_i')&=&\frac{1}{2\U}\left\{-\lambda_i'+\T\frac{4N_c \left( \T^2-\U^2 \right)-2\lambda_i'
\left( 2\T-\U \right)}{\left[ N_c \left( \T+\U \right)-\lambda_i' \right] \left(
\lambda_i'-2N_c\U \right)}\right\} ,
\nonumber \\
x_2(\lambda_i')&=&-\frac{1}{2\T}\left[ \lambda_i'+2\U x_1(\lambda_i') \right] ,
\nonumber \\
x_3(\lambda_i')&=&\frac{1}{4\U\T^2}\left\{\left[ \lambda_i'+2\U x_1(\lambda_i') \right] \left[
\left( \lambda_i'-2N_c\U \right) \left( \lambda_i'-N_c\U \right)
-2\T \left( \T-\U \right) \right] \right.
\nonumber \\ && \quad \quad \quad
\left. -2\T \left( \T-\U \right) \left( \lambda_i'-2N_c\U \right) \right\} ,
\nonumber \\ 
x_4(\lambda_i')&=&\frac{1}{\U}\left[ \left(N_c\T-\lambda_i' \right) x_1(\lambda_i')+\T-\U\right] ,
\nonumber \\
x_6(\lambda_i')&=&\frac{1}{\T}\left\{\frac{\left( \lambda_i'-N_c\U \right) \left[
\lambda_i'+2\U x_1(\lambda_i') \right]}{2\T}+\U-\T\right\}.
\end{eqnarray}

Working with $N_c=3$, the eigenvalues simplify to
\beqa
\lambda _1&=&\lambda_4=3 \frac{\alpha _s}{\pi} \T ,
\nonumber \\ 
\lambda _2&=&\lambda_5=3 \frac{\alpha _s}{\pi} \U ,
\nonumber \\ 
\lambda _3&=&\lambda_6=3 \frac{\alpha _s}{\pi} \left( \T+\U \right) , 
\nonumber \\ 
\lambda _7&=&2 \frac{\alpha _s}{\pi} (\T+\U) ,
\nonumber \\ 
\lambda _{8,9}&=&\frac{\alpha _s}{\pi}\left[2\T+2\U\pm4\sqrt{\T^2+\U^2-\T\U}\right].
\label{eigvalcol3}
\eeqa
For $N_c=3$, the eigenvectors
$e_4^{(6)}$,  $e_5^{(6)}$ and 
$e_6^{(6)}$ become (compare  Eq.\ (\ref{456eigvc}))
\beq
e_4^{(6)}=\left(\begin{array}{c}
-\frac{7\T}{6\U} \vspace{2mm} \\ 
-\frac{1}{3} \vspace{2mm}\\
\frac{\U-\T}{3\U} \vspace{2mm}\\
\frac{\T-\U}{\U} \vspace{2mm}\\
1   \vspace{2mm}\\
0
\end{array}\right) , \; \; \; 
e_5^{(6)}=\left(\begin{array}{c}
\frac{1}{3} \frac{\T}{\T-\U}  \vspace{2mm}\\ 
\frac{7\U}{6\left( \T-\U \right)}  \vspace{2mm}\\
-\frac{1}{3}  \vspace{2mm}\\
0 \vspace{2mm} \\
-\frac{\T}{\T-\U}    \vspace{2mm}\\
1
\end{array}\right) , \; \; \; 
e_6^{(6)}=\left(\begin{array}{c}
\frac{\T}{3\U} \vspace{2mm}\\ 
-\frac{1}{3}\\
\frac{7\left(\T-\U\right)}{6\U} \vspace{2mm}\\
-\frac{\T}{\U} \vspace{2mm}\\
0   \vspace{2mm}\\
1
\end{array}\right),
\eeq
while $e_7^{(6)}$,  $e_8^{(6)}$ and 
$e_9^{(6)}$ reduce to
\beq
e_7^{(6)}=\left(\begin{array}{c}
-1 \\ 
-1\\
-1 \\
1 \\
1   \\
1
\end{array}\right) , \; \; \;
e_{i}=\left(\begin{array}{c}
a_1(\lambda_i') \\ 
a_2(\lambda_i') \\
a_3(\lambda_i') \\
a_4(\lambda_i') \\
1   \\
a_6(\lambda_i')
\end{array}\right), \; i=8,9 \, .
\eeq
Here for brevity  we have introduced  the quantities
\beqa
a_1(\lambda_i')&=&\frac{-1}{6 \U K}\left[20 (\T^3-\U^3)-52 \T \U (\T-\U)
+(10\T^2+5 \U^2-9\T\U) \lambda_i'\right] ,
\nonumber \\
a_2(\lambda_i')&=&a_1(\lambda_i',\T \leftrightarrow \U) ,
\nonumber \\
a_3(\lambda_i')&=&\frac{1}{6 \T \U K}\{-36 \T \U(\T^2+\U^2) +32 \T^2 \U^2 +20 (\T^4+\U^4) 
\nonumber \\ &&
+[10 (\T^3 + \U^3) +5 \T \U (\T+\U)] \lambda_i'\} ,
\nonumber \\
a_4(\lambda_i')&=&\frac{1}{2 \U^2 K} [10 \U^4 +20 \T^4 -54 \T^3 \U +58 \T^2 \U^2 -34 \T \U^3
\nonumber \\ &&
+(10 \T^2  +7  \U^2  -7 \T \U) \T \lambda_i'] ,
\nonumber \\
a_6(\lambda_i')&=&a_4(\lambda_i',\T \leftrightarrow  \U) ,
\eeqa
where the $\lambda_i'$'s are obtained from the $\lambda_i$'s of (\ref{eigvalcol3})
according to Eq.\ (\ref{redeival}). The quantity $K$ is defined as
\beq
K\equiv 5 (\T^2+\U^2)- 6 \U \T.
\label{kappadef}
\eeq
Using the formalism of subsection \ref{chabasexp}, we can write the
matrix ${\left( R^{(\a)} \right)}^{-1}$, having the eigenvectors 
of the anomalous dimension matrix in its columns. We have verified 
that $R^{(\a)}{ \Gamma_{S'} \left( R^{(\a)} \right)}^{-1}$ is a diagonal
matrix with the eigenvalues of Eq.\ (\ref{eigvalcol3}).

\subsection{Color projections for $gg \rightarrow gg$}
\label{subsec-projbas}

So far we have been working with the sets of color structures in Eqs.\ (\ref{basgg1}) 
and (\ref{basgg2}). These sets are both overcomplete, since the $c_I$'s  are  not linearly
independent \cite{Macfar}, as we will show below.
In this section we would like to rewrite the set in Eq.\ (\ref{basgg2}) in terms of $SU(3)$ 
tensors (we keep $N_c=3$), and to elucidate the group structure of 
$t$-channel color exchange for the product of
Wilson lines that generates the same noncollinear soft radiation as $gg \rightarrow gg$.

To relate the basis of traces to color
exchange, we first recall the product formula for 
generators of $SU(N_c)$ in the fundamental representation,
\beq
T_{F}^i T_{F}^j=\frac{1}{6} \delta_{ij}{{1}}+\frac{1}{2}\left( d_{ijk}
+if_{ijk} \right) T_{F}^k \, .
\label{eq:cons}
\eeq
Eq.\ (\ref{eq:cons}) enables us to rewrite the set of color structures in Eq.\ (\ref{basgg2}) 
in terms of the tensors $f$ and $d$,
\beqa
{c'}_1={\rm{Tr}}\left( T_{F}^{\e_A}T_{F}^{\e_B}T_{F}^{\e_2}T_{F}^{\e_1} \right)-{\rm{Tr}}
\left( T_{F}^{\e_A}T_{F}^{\e_1}T_{F}^{\e_2}T_{F}^{\e_B} \right)&=&\frac{i}{4}\left[f_{\e_A \e_B l}
d_{\e_1 \e_2 l}-d_{\e_A \e_B l}f_{\e_1 \e_2 l}\right], \nonumber \\
{c'}_2={\rm{Tr}}\left( T_{F}^{\e_A}T_{F}^{\e_B}T_{F}^{\e_1}T_{F}^{\e_2} \right)-{\rm{Tr}}
\left( T_{F}^{\e_A}T_{F}^{\e_2}T_{F}^{\e_1}T_{F}^{\e_B} \right)&=&\frac{i}{4}\left[f_{\e_A \e_B l}
d_{\e_1 \e_2 l}+d_{\e_A \e_B l}f_{\e_1 \e_2 l}\right], \nonumber \\
{c'}_3={\rm{Tr}}\left( T_{F}^{\e_A}T_{F}^{\e_1}T_{F}^{\e_B}T_{F}^{\e_2} \right)-{\rm{Tr}}
\left( T_{F}^{\e_A}T_{F}^{\e_2}T_{F}^{\e_B}T_{F}^{\e_1} \right)&=&\frac{i}{4}
\left[f_{\e_A \e_1 l}d_{\e_B \e_2 l}+d_{\e_A \e_1 l}f_{\e_B \e_2 l}\right],\nonumber \\
{c'}_4={\rm{Tr}}\left( T_{F}^{\e_A}T_{F}^{\e_B}T_{F}^{\e_2}T_{F}^{\e_1} \right)+{\rm{Tr}}
\left( T_{F}^{\e_A}T_{F}^{\e_1}T_{F}^{\e_2}T_{F}^{\e_B} \right)&=&\frac{1}{6}\delta_{\e_A\e_1}
\delta_{\e_B\e_2}+\frac{1}{4}\left[d_{\e_A \e_1 l}d_{\e_B \e_2 l} \right. 
\nonumber \\&& \left. \quad \quad \quad \quad 
+f_{\e_A \e_1 l}f_{\e_B \e_2 l}\right],\nonumber \\
{c'}_5={\rm{Tr}}\left( T_{F}^{\e_A}T_{F}^{\e_B}T_{F}^{\e_1}T_{F}^{\e_2} \right)+{\rm{Tr}}
\left( T_{F}^{\e_A}T_{F}^{\e_2}T_{F}^{\e_1}T_{F}^{\e_B} \right)&=&\frac{1}{6}\delta_{\e_A\e_B}
\delta_{\e_1\e_2}+\frac{1}{4}\left[d_{\e_A \e_B l}d_{\e_1 \e_2 l} \right. 
\nonumber \\&& \left.  \quad \quad \quad \quad
-f_{\e_A \e_B l}f_{\e_1 \e_2 l}\right],\nonumber \\
{c'}_6={\rm{Tr}}\left( T_{F}^{\e_A}T_{F}^{\e_1}T_{F}^{\e_B}T_{F}^{\e_2} \right)+{\rm{Tr}}
\left( T_{F}^{\e_A}T_{F}^{\e_2}T_{F}^{\e_B}T_{F}^{\e_1} \right)&=&\frac{1}{6}\delta_{\e_A\e_1}
\delta_{\e_B\e_2}+\frac{1}{4}\left[d_{\e_A \e_1 l}d_{\e_B \e_2 l} \right. 
\nonumber \\ && \left. \quad \quad \quad \quad
-f_{\e_A \e_1 l}f_{\e_B \e_2 l}\right], \nonumber \\
{c'}_7=\frac{1}{4}\delta_{\e_A\e_1}\delta_{\e_B\e_2} \, , \nonumber \\
{c'}_8=\frac{1}{4}\delta_{\e_A\e_B}\delta_{\e_1\e_2} \, , \nonumber \\
{c'}_9=\frac{1}{4}\delta_{\e_A\e_2}\delta_{\e_B\e_1} \, . 
\label{eq:dfbasis}
\eeqa
We now go on to  discuss color in the $t$-channel.

The color content of a set of two gluons in $SU(3)$ is described by the direct product
$8 \otimes 8$, which can be decomposed into irreducible representations in the standard way, as
\beq
8 \otimes 8=1+8_S+8_A+10+{\overline {10}}+27.
\label{col88dec}
\eeq  
In Ref.\ \cite{Bartels} we can find the representation in terms of $SU(3)$ tensors
of the projectors performing the above decomposition. In our notation, 
$t$-channel projectors are given by
\beqa
P_1(\e_A,\e_B;\e_1,\e_2)&=&\frac{1}{8}\delta_{\e_A \e_1} \delta_{\e_B \e_2} \, ,
\nonumber \\ 
P_{8_S}(\e_A,\e_B;\e_1,\e_2)&=&\frac{3}{5} d_{\e_A\e_1c} d_{\e_B\e_2c} \, ,
\nonumber \\ 
P_{8_A}(\e_A,\e_B;\e_1,\e_2)&=&\frac{1}{3} f_{\e_A\e_1c} f_{\e_B\e_2c} \, ,
\nonumber \\ 
P_{10+{\overline {10}}}(\e_A,\e_B;\e_1,\e_2)&=&
\frac{1}{2}(\delta_{\e_A \e_B} \delta_{\e_1 \e_2}
-\delta_{\e_A \e_2} \delta_{\e_B \e_1})
-\frac{1}{3} f_{\e_A\e_1c} f_{\e_B\e_2c} \, ,
\nonumber \\ 
P_{27}(\e_A,\e_B;\e_1,\e_2)&=&\frac{1}{2}(\delta_{\e_A \e_B} \delta_{\e_1 \e_2}
+\delta_{\e_A \e_2} \delta_{\e_B \e_1})
-\frac{1}{8}\delta_{\e_A \e_1} \delta_{\e_B \e_2}
\nonumber \\ && 
-\frac{3}{5} d_{\e_A\e_1c} d_{\e_B\e_2c} \, .
\label{88proj}
\eeqa
From a comparison of Eq. (\ref{eq:dfbasis}) and Eq.\ (\ref{88proj}), we see that the first 
three elements of the basis (\ref{eq:dfbasis}), depending only on the ``mixed'' combinations $fd$, $df$,
cannot be expressed in terms of these projectors. 
On the other hand, it
is easy to find relations for the remaining basis tensors, 
\beqa
{c'}_4&=&\frac{4}{3}P_1+\frac{5}{12}P_{8_S}+\frac{3}{4}P_{8_A} \, ,
\nonumber \\
{c'}_5&=&-\frac{1}{6}P_1-\frac{1}{3}P_{8_S}+\frac{1}{2}P_{27} \, ,
\nonumber \\
{c'}_6&=&\frac{4}{3}P_1+\frac{5}{12}P_{8_S}-\frac{3}{4}P_{8_A} \, ,
\nonumber \\
{c'}_7&=&2P_1 \, , \nonumber \\
{c'}_8&=&\frac{1}{4}\left( P_1+P_{8_A}+P_{8_S}+P_{10 \oplus \overline{10}}
+P_{27} \right) \, , \nonumber \\
{c'}_9&=&\frac{1}{4}\left( P_1-P_{8_A}+P_{8_S}-P_{10 \oplus \overline{10}}
+P_{27} \right) \, .
\label{eq:funcproj}
\eeqa
Evidently, the six tensors $c'_I$, $i=4\dots9$ may be replaced by only five
$t$-channel projectors.  This is because, as pointed out above,
the original basis of Eq.\ (\ref{basgg1}) is overcomplete, and indeed,
the basis elements of Eq.\ (\ref{basgg2}) satisfy the relation 
\beq
{c'}_4+{c'}_5+{c'}_6={c'}_7+{c'}_8+{c'}_9 \, .
\label{eq:reduce2}
\eeq
In terms of the invariant tensor $d_{ijk}$, related to the
$c_I'$ in Eq.\ (\ref{eq:dfbasis}), this is equivalent to the identity \cite{Macfar}
\beq
d_{i\ell m}d_{mjk}+d_{j\ell m}d_{imk}+d_{k\ell m}d_{ijm}
=
{1\over 3}\left( \delta_{ij}\delta_{k\ell}+\delta_{jk}\delta_{i\ell}+\delta_{ik}\delta_{j\ell}\right)\, .
\label{ddidentity}
\eeq
It may be worth noting that the Jacobi identity, as well as a related identity
involving products of $f$'s and $d$'s in Ref.\ \cite{Macfar}, is simply an expression
of the cyclicity of traces in the original basis, Eq.\ (\ref{basgg1}),
and hence does not result in a further reduction of its dimension.

Our new basis is thus given by the eight color structures
\beq
\left\{ {c'}_1,{c'}_2,{c'}_3,P_1,P_{8_S},P_{8_A},P_{10 \oplus 
\overline{10}},P_{27}\right\}\, .
\label{eq:baslast}
\eeq
In this basis, the soft anomalous dimension matrix becomes
\beq
\Gamma_{S'}=\left(\begin{array}{cc}
            \Gamma_{3 \times 3} & 0_{3 \times 5} \\
              0_{5 \times 3}      & \Gamma_{5 \times 5}
\end{array} \right),
\label{eq:gambloproj}
\eeq
with $\Gamma_{3 \times 3}$ given by Eq.\ (\ref{eq:subbloc1}) 
with $N_c=3$, and with $\Gamma_{5 \times 5}$ given by
\beq
\Gamma_{5 \times 5}=\frac{\alpha_s}{\pi}\left(\begin{array}{ccccc}
6\T & 0 & -6\U & 0 & 0 \\
0  & 3\T+\frac{3\U}{2} & -\frac{3\U}{2} & -3\U & 0 \\ \vspace{2mm}
-\frac{3\U}{4} & -\frac{3\U}{2} &3\T+\frac{3\U}{2} & 0 & -\frac{9\U}{4} \\ 
\vspace{2mm}
0 & -\frac{6\U}{5} & 0 & 3\U & -\frac{9\U}{5} \\ \vspace{2mm} 
0 & 0 &-\frac{2\U}{3} &-\frac{4\U}{3} & -2\T+4\U 
\end{array} \right).
\label{eq:Gamma55}
\eeq
From Eqs.\ (\ref{eq:subbloc1}), (\ref{eq:gambloproj})  and  (\ref{eq:Gamma55}), we 
see that the anomalous dimension
matrix is now  diagonal in its $\T$ dependence. 

We have computed the eigenvalues of the matrix above, to check that they are the same as in  
Eq.\ (\ref{eigvalcol3}) (we will have one fewer, due to the dimensional reduction of our matrix),
finding, indeed,
\beqa
\lambda_4&=&\lambda_1=3 \frac{\alpha_s}{\pi} \T \, ,\nonumber\\
\lambda_5&=&\lambda_2=3 \frac{\alpha_s}{\pi} \U \, ,\nonumber\\
\lambda_6&=&\lambda_3=3 \frac{\alpha_s}{\pi} (\T+\U) \, ,\nonumber\\
\lambda_{7}&=&2 \frac{\alpha_s}{\pi} \left[\T+\U - 2\sqrt{\T^2-\T\U+\U^2}\right] \, ,\nonumber\\
\lambda_{8}&=&2 \frac{\alpha_s}{\pi} \left[\T+\U + 2\sqrt{\T^2-\T\U+\U^2}\right] \, . 
\label{eq:eigvaproj}
\eeqa
The eigenvectors have the general form
\beq
e_i=\left(\begin{array}{c}
      e_i^{(3)} \\ 
       0^{(5)} 
\end{array}\right), \; i=1,2,3 \, , \; \; \; 
e_i=\left(\begin{array}{c}
     0^{(3)} \\ 
     e_i^{(5)} 
\end{array}\right), \; i=4 \ldots 8 \, ,
\label{eq:35parteeigvproj}
\eeq
where the $e_i^{(3)}$ are given explicitly in Eq.\ (\ref{eq:3parteigv});
$e_4^{(5)}$,  $e_5^{(5)}$ and 
$e_6^{(5)}$ are given by
\beq
e_4^{(5)}=\left(\begin{array}{c}
      -15 \vspace{2mm} \\ 
     6-\frac{15}{2}\frac{\T}{\U} \vspace{2mm} \\
      -\frac{15}{2}\frac{\T}{\U} \vspace{2mm} \\ 
        3 \vspace{2mm} \\ 
        1 \end{array} \right), \; \; \;
e_5^{(5)}=\left(\begin{array}{c}
      0   \vspace{2mm} \\
      -\frac{3}{2} \vspace{2mm} \\
       0  \vspace{2mm} \\
      \frac{3}{4}-\frac{3}{2}\frac{\T}{\U} \vspace{2mm}  \\
       1 \end{array} \right), \; \; \; 
e_6^{(5)}=\left(\begin{array}{c}
      -15  \vspace{2mm} \\
     -\frac{3}{2}+\frac{15}{2}\frac{\T}{\U} \vspace{2mm}  \\ 
      \frac{15}{2}-\frac{15}{2}\frac{\T}{\U} \vspace{2mm} \\ 
        -3 \vspace{2mm} \\ 
        1 \end{array} \right);
\label{eq:v1v2v3proj}
\eeq
$e_7^{(5)}$ and $e_8^{(5)}$ have a rather complicated form, which can be simplified by
the reparameterization,
\beqa
e_i^{(5)}&=&  \left(\begin{array}{c}
        b_1(\lambda_i') \\
        b_2(\lambda_i') \\ 
        b_3(\lambda_i') \\
        b_4(\lambda_i') \\
        1 
\end{array} \right), \; i=7,8    
\eeqa
where $\lambda_i'$ was defined in Eq.\ (\ref{redeival}), and where the $b_i$'s are given by 
\beqa
b_1(\lambda_i')&=&\frac{3}{\U^2 K'} [80 \T^4+103 \U^4-280 \U \T^3 -300 \T \U^3 +404 \T^2 \U^2
\nonumber \\ && \quad \quad
+(40 \T^3-16 \U^3 -60 \T^2 \U +52 \T\U^2)\lambda_i'] \, ,
\nonumber \\
b_2(\lambda_i')&=&\frac{3}{2K'}[20 \T^2-50 \U\T +44 \U^2+(10 \T-5 \U)\lambda_i'] \, ,
\nonumber \\
b_3(\lambda_i')&=&-\frac{3}{2 \U K'}[40\T^3-64\U^3-120 \T^2 \U+130 \T \U^2
\nonumber \\ && \quad \quad
+(20 \T^2+13 \U^2-20 \T\U) \lambda_i'] \, ,
\nonumber \\
b_4(\lambda_i')&=&\frac{3\U}{K'}(2\T+5\U-2\lambda_i') \, ,
\eeqa
$K'$ being the auxiliary function
\beq
K'=20 \T^2-20 \U\T+21 \U^2 \, .
\eeq

As $T\rightarrow-\infty$, 
the eigenvectors $e_i$ approach simple linear combinations of the color
tensors of Eq.\ (\ref{eq:baslast}).  In particular, $e_7$
becomes proportional to  the color singlet vector $(0,0,0,1,0,0,0,0)$
in Eq.\ (\ref{eq:baslast}), with an eigenvalue that decreases as $6T$.
This is the most negative eigenvalue, so that, once again, color singlet
exchange dominates in the product of hard and soft functions, Eq.\ (\ref{softsoleigbas}),
for gluon-gluon scattering.  
Of the remaining eigenvalues, the color structures of the first three
are simple products of the $f$ and $d$ tensors for all angles (see Eq.\ (\ref{eq:dfbasis})).
Next, from Eq.\ (\ref{eq:v1v2v3proj}), we see that $e_4$ and $e_6$ become degenerate and approach
linear combinations of symmetric and antisymmetric octet projectors,
while $e_5$ corresponds to $10\oplus \overline{10}$ exchange in the forward limit.
Finally, $e_8$ approaches pure $27$ exchange in the forward limit,
with the {\it largest}, and hence most suppressed, eigenvalue, $-2T$.
In fact, suppression increases with the dimension of the color
representation exchanged.  This suggests that
radiation by accelerated charges in a nonabelian gauge theory increases
with the dimension of the color exchange, not only from singlet
to octet, but also to higher representations. 

\section{Conclusions}

The anomalous dimension matrix for soft radiation in gluon-gluon scattering
completes the set of one-loop calculations that we set out to analyze.
An interesting regularity of our results,
relevant to near-forward scattering,
is that for each flavor combination, logarithms of
$t/s$ appear only in diagonal matrix elements in those bases 
that describe definite
color exchange in the $t$-channel.  This had been shown
previously for the cases of quark-quark and quark-antiquark scattering
at one loop \cite{SotiSt,GK} and two loops  \cite{KK}.
One application of the anomalous dimension matrices and their eigenvalues is for the
resummation of threshold singularities to next-to-leading logarithm in jet cross sections \cite{KOS1}
at fixed rapidities.  This was the original motivation for our study.
We are hopeful, however, that the formalism of resumming coherent noncollinear
soft gluon radiation in terms of 
color mixing will have a variety of other uses as well.

\section*{Acknowledgements}

This work was supported in part by the National Science Foundation,
under grant PHY9309888 and by the PPARC under grant GR/K54601.
We wish to thank Eric Laenen,
Jack Smith, and Ramona Vogt for many helpful conversations.

\appendix
\section{Feynman rules for eikonal diagrams in axial gauge}
\label{app-feyn}

In this Appendix we summarize the  Feynman rules we have
used in the evaluation of the eikonal diagrams of Fig.\ \ref{fig_eiko}.
We refer to Fig.\ \ref{fig_Feyn}. 

The propagator for a quark, antiquark or gluon eikonal line (Fig.\ \ref{fig_Feyn}) is given by
\beq
\frac{i}{\delta v \cdot q + i\epsilon},
\label{qprop}
\eeq
where $\delta=+1(-1)$ for the momentum $q$ flowing 
in the same (opposite) direction as the dimensionless vector $v$.
The interaction vertex for a quark or antiquark eikonal line (Fig.\ \ref{fig_Feyn}a, b)
is 
\beq
-ig_s {T_F^c}_{ba} \Delta v^{\mu},
\eeq 
with $\Delta=+1(-1)$
for a quark (antiquark).

For the gluon eikonal vertex (Fig.\ \ref{fig_Feyn}c), we may take
\beq
-g_s f^{abc} \Delta v_{\mu},
\eeq
where we agree to read the color indexes $a,b,c$ anticlockwise, and where 
$\Delta=+1(-1)$ for the gluon located below (above) the eikonal line.

Finally
in our calculations we 
employ the general axial gauge gluon propagator,
\begin{equation}
D^{\mu \nu}(k)=\frac{-i}{k^2+i\epsilon} N^{\mu \nu}(k), \quad
N^{\mu \nu}(k)=g^{\mu \nu}-\frac{n^{\mu}k^{\nu}+k^{\mu}n^{\nu}}{n \cdot k}
+n^2\frac{k^{\mu}k^{\nu}}{(n \cdot k)^2},
\end{equation}
with $n^{\mu}$ the axial gauge-fixing vector.

\section{Evaluation of one-loop eikonal vertex corrections}
\label{app-integ}

\begin{figure}
\centerline{\epsffile{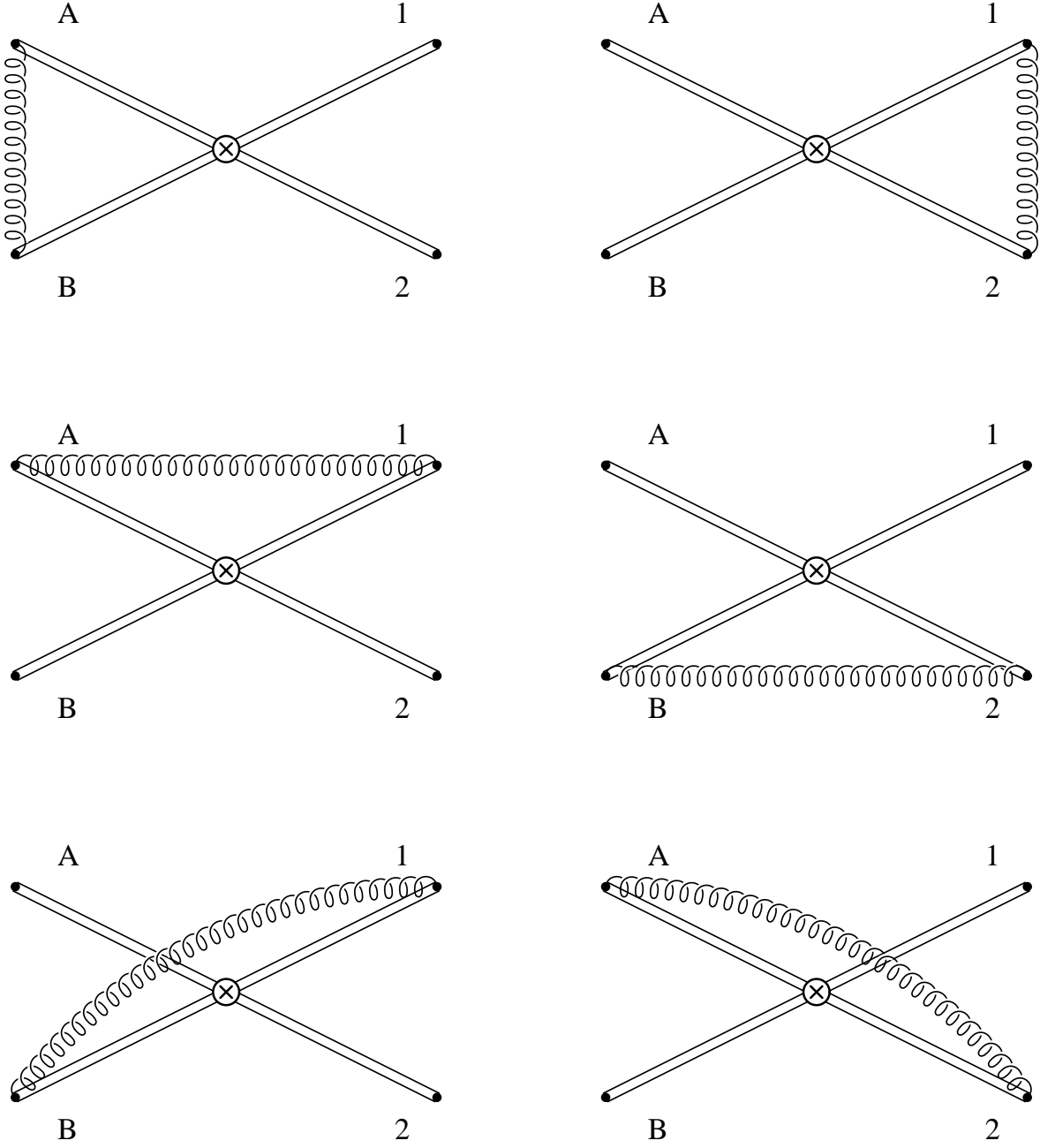}}
\caption{Eikonal vertex corrections for partonic processes
contributing to the soft anomalous dimension matrices.}
\label{fig_eiko}
\end{figure}

We refer to Fig.\ \ref{fig_eiko}, where we have represented the 
UV divergent $O(\alpha_s)$ contributions to 
the eikonal vertex  $W_{I}^{(\a)}$.
The counterterms for $W_I^{(\a)}$ are  ultraviolet divergent
coefficients, depending on the external momenta and on the axial gauge fixing vector,
times our basis color tensors, $c_I^{(\a)}$.

At Born level the eikonal vertices describe only color flow, and the 
relation 
\beq
W_{I, \, \rm{Born}}^{(\a)}=c_I^{(\a)}
\eeq
holds.

The one-loop corrections can be written in matrix notation as
\beq
W_{\rm{1 loop}}^{(\a), T} \, = \, c^{(\a), T} \, Z_S \,= \, W_{\rm{Born}}^{(\a), T} \, Z_S\, ,
\label{oneloopcorr}
\eeq
where the superscript $T$ indicates transpose. The dimension of the vectors (and therefore
the rank
of the matrix of renormalization constants) depends on the specific partonic process, as
discussed in detail above.

We denote the kinematic part of the one-loop vertex correction to $W_{I}^{(\a)}$,  
with the virtual gluon linking lines $v_{i}$ and $v_{j}$,
as $\omega_{ij}(\delta_{i}v_{i},\delta_{j}v_{j},\Delta_{i},\Delta_{j})$,
with $\delta_i$'s and $\Delta_i$'s defined in Appendix \ref{app-feyn}.
Its expression is
\beqa   
\omega_{ij}(\delta_{i}v_{i},\delta_{j}v_{j},\Delta_{i},\Delta_{j})&=&
{g}_{s}^2\int\frac{d^nq}{(2\pi)^n}\frac{-i}{q^2+i\epsilon}
\left\{\frac{\Delta_{i} \: \Delta_{j} \:v_{i}{\cdot}v_{j}}{(\delta_{i}v_{i}{\cdot}q+i\epsilon)
(\delta_{j}v_{j}{\cdot}q+i\epsilon)}\right. \nonumber\\
\hspace{-35mm} & &\left.-\frac{\Delta_{i}v_{i}{\cdot}n}{(\delta_{i}v_{i}{\cdot}q+i\epsilon)}
\frac{P}{(n{\cdot}q)}-\frac{\Delta_{j}v_{j}{\cdot}n}{(\delta_{j}v_{j}{\cdot}q+i\epsilon)}
\frac{P}{(n{\cdot}q)}+n^2\frac{P}{(n{\cdot}q)^2}\right\}, \nonumber \\
& &        \label{eq:omega}
\eeqa
where $P$ stands for principal value,
\beqa
\frac{P}{(q \cdot n)^{\beta}}=\frac{1}{2}\left(\frac{1}{(q \cdot n+i\epsilon)
^{\beta}}+(-1)^{\beta}\frac{1}{(-q \cdot n+i\epsilon)^{\beta}}\right).
\eeqa
We rewrite (\ref{eq:omega}) as
\beqa   
\omega_{ij}(\delta_{i}v_{i},\delta_{j}v_{j},\Delta_{i},\Delta_{j})&=& {\cal S}_{ij} 
\left[I_1(\delta_i v_i, \delta_j v_j)
-\frac{1}{2}I_2(\delta_i v_i, n)-\frac{1}{2}I_2(\delta_i v_i, -n)\right.
\nonumber \\ && 
\left.-\frac{1}{2}I_3(\delta_j v_j, n)-\frac{1}{2}I_3(\delta_j v_j, -n)
+I_4(n^2)\right] \, ,
\label{eq:omsum}
\end{eqnarray}
where ${\cal S}_{ij}$ is an  overall sign
\beq
{\cal S}_{ij}=\Delta_i \: \Delta_j \: \delta_i \: \delta_j.
\eeq
The ultraviolet poles of the integrals are   \cite{BottsSt,SotiSt}
\beqa
I_1^{UV \;pole}&=&\frac{\alpha_s}{\pi}
\left\{\frac{2}{\epsilon^2}-\frac{1}{\epsilon}
\left[\gamma+\ln\left(\delta_i\delta_j\; \frac{v_i \cdot v_j}{2}\right)
-\ln(4\pi) \right]\right\} \, , 
\nonumber\\ 
I_2^{UV \;pole}&=&\frac{\alpha_s}{2\pi}
\left\{\frac{2}{\epsilon^2}-\frac{1}{\epsilon}
\left[\gamma+\ln(\nu_i)-\ln(4\pi)\right]\right\} \, , 
\nonumber\\
I_3^{UV \;pole}&=&\frac{\alpha_s}{2\pi}
\left\{\frac{2}{\epsilon^2}-\frac{1}{\epsilon}
\left[\gamma+\ln(\nu_j)-\ln(4\pi)\right]\right\} \, , 
\nonumber\\
I_4^{UV \;pole}&=&-\frac{\alpha_s}{\pi}\frac{1}{\epsilon}\, ,
\label{eq:UVpoles} 
\eeqa
where all the gauge dependence is through the variable
\beq
\nu_i=\frac{(v_{i}{\cdot}n)^2}{|n|^2}.
\label{eq:nu}
\eeq
The double poles cancel in (\ref{eq:omsum}), giving
\beq
\omega_{ij}(\delta_{i}v_{i},\delta_{j}v_{j},\Delta_{i},\Delta_{j})=- {\cal S}_{ij} \: \frac{\alpha_{s}}{\pi\epsilon}\left[\ln(\frac{\delta_{i} \: \delta_{j} \: v_{i}{\cdot}v_{j}}{2})-\frac{1}{2}\ln(\nu_{i}\nu_{j})+1\right].
\label{eq:omfin}
\eeq
In order to obtain contributions to the different entries of the matrix of renormalization constants,
the above expression has still to be multiplied by the color decomposition of its corresponding 
diagram into the basis color structures \cite{BottsSt,KS,Thesis}.

\newpage

\begin{figure}
\centerline{\epsffile{fig_coldec.epsi}}
\caption{Graphical representation for the color decompositions in
Eq.\ ({\protect \ref{breaktrace}}).}
\label{fig_coldec}
\end{figure}

\begin{figure}
\centerline{\epsffile{fig_Feyn.epsi}}
\caption{Illustration of the eikonal Feynman rules for a quark (a), antiquark (b) and gluon (c) eikonal line.}
\label{fig_Feyn}
\end{figure}


\begin{thebibliography}{99}

\bibitem{CSS1} G.T.\ Bodwin, Phys.\ Rev.\ D31, 2616 (1985); E.\ D34, 3932 (1986);
J.C.\ Collins, D.E.\ Soper and G.\ Sterman, Nucl.\ Phys.\ B261, 104 (1985).

\bibitem{CSS88} J.C.\ Collins, D.E.\ Soper and G.\ Sterman, Nucl.\ Phys.\ B308, 833 (1988).

\bibitem{CSSrv} J.C.\ Collins, D.E.\ Soper and G.\ Sterman,
in {\it Perturbative Quantum Chromodynamics},
ed.\ A. H.\ Mueller (World Scientific, Singapore, 1989), p. 1.

\bibitem{CoSo81} J.C.\ Collins and D.E.\ Soper, Nucl.\ Phys.\ B193, 381 (1981).

\bibitem{DYqt1}      Y.L.\ Dokshitzer, D.I.\ Dyakonov, and S.I.\ Troyan,
     Phys.\ Lett.\ 79B, 269 (1978);
   G.\ Parisi and R.\ Petronzio, Nucl.\ Phys.\ B154, 427 (1979).

\bibitem{DYqt2} J.C.\ Collins, D.E.\ Soper, and G.\ Sterman,
     Nucl.\ Phys.\ B250, 199 (1985);
     C.T.H.\ Davies and W.J.\ Stirling, Nucl.\ Phys.\ B244, 337 (1984);
  C.T.H.\ Davies, B.R.\ Webber, and W.J.\ Stirling, Nucl.\ Phys.\ B256,
      413 (1985);
   P.B.\ Arnold and R.P.\ Kauffman, Nucl.\ Phys.\ B349, 381 (1991);
   G.\ Ladinsky and C.-P.\ Yuan, Phys.\ Rev.\ D 50, 4239 (1994).

\bibitem{BottsSt} 
J.\ Botts and G.\ Sterman, Nucl.\ Phys.\ {B325}, 62 (1989).

\bibitem{SotiSt}
M.G.\ Sotiropoulos and G.\ Sterman, Nucl.\ Phys.\ {B419}, 59 (1994).

\bibitem{GK} G.P.\ Korchemsky, Phys.\ Lett.\ B {325}, 459 (1994).

\bibitem{KK} I.A.\ Korchemskaya and G. P.\ Korchemsky,
Nucl.\ Phys.\ {B437}, 127 (1995).

\bibitem{CoCoh} 
Y.L.\ Dokshitzer, D.I.\ Dyakonov and S.I.\ Troyan, Phys.\ Rep.\ 58, 269 (1980);
A.H.\ Mueller, Phys.\ Lett.\ 104B, 161 (1981);
A. Bassetto, M. Ciafaloni, G. Marchesini and A.H.\ Mueller, Nucl.\ Phys.\ B207, 189 (1982); 
Y.L.\ Dokshitzer,  V.A.\ Khoze, S.I.\ Troyan and A.H.\ Mueller, Rev.\ Mod.\ Phys.\ 60, 373 (1988);
G.\ Marchesini and B.R.\ Webber, Nucl.\ Phys.\ B310, 461 (1988); 
Y.L.\ Dokshitzer, V.A.\ Khoze and
S.I. Troyan, in {\it Perturbative Quantum Chromodynamics}, ed. A.H. Mueller
(World Scientific, Singapore, 1989), p.\ 241;
Y.L.\ Dokshitzer, V.A.\ Khoze, G.\ Marchesini and B.R.\ Webber,
Phys.\ Lett.\ B245, 243 (1990).

\bibitem{St87} G.\ Sterman, Nucl.\ Phys.\ {B281}, 310 (1987).

\bibitem{CT} S.\ Catani and L.\ Trentadue,
Nucl.\ Phys.\ {B327}, 323 (1989); B353, 183 (1991);
L.\ Magnea, Nucl.\ Phys.\ B349, 703 (1991).

\bibitem{LSvNKV} E.\ Laenen, J.\ Smith, and W.L.\ van Neerven,
Nucl. Phys. B369, 543 (1992);
Phys. Lett. B321, 254 (1994); N.\ Kidonakis and J.\ Smith,
Phys.\ Rev.\ D51, 6092 (1995); Mod.\ Phys.\ Lett.\ A11, 587 (1996).

\bibitem{BC} E.L.\ Berger and H.\ Contopanagos, Phys. Lett. B361,
115 (1995); Phys. Rev. D54, 3085 (1996); {\it ibid.} D57, 253 (1998).

\bibitem{CMNT} S.\ Catani, M.L.\ Mangano, P.\ Nason, and L.\ Trentadue,
Phys. Lett. B378, 329 (1996); Nucl. Phys. {B478}, 273 (1996).

\bibitem{KS} N.\ Kidonakis and G.\ Sterman, Phys.\ Lett.\ B387, 867 (1996); 
Nucl.\ Phys.\ B505, 321 (1997); in proceedings of
{\it Deep Inelastic
Scattering and QCD, 5th International Workshop}, ed.\ J.\ Repond and
D.\ Krakauer (AIP Conf.\ Proc.\ No.\ 407, American Institute of
Physics, Woodbury, NY, 1997), p.\ 1035, hep-ph/9708353.

\bibitem{Thesis} N. Kidonakis, SUNY at Stony Brook Ph.D. Thesis (1996),
hep-ph/9606474.

\bibitem{NKJSRV} N.\ Kidonakis, J.\ Smith and R.\ Vogt,
Phys.\ Rev.\ D56, 1553 (1997); N.\ Kidonakis, contribution presented 
at the QCD 97 Euroconference, Montpellier, July 3-9, 1997, hep-ph/9708439;
N.\ Kidonakis and R.\ Vogt, in preparation.

\bibitem{BCMN} R.\ Bonciani, S.\ Catani, M.L.\ Mangano and P.\ Nason,
hep-ph/9801375.

\bibitem{KOS1} N.\ Kidonakis, G.\ Oderda and G.\ Sterman, hep-ph/9801268.

\bibitem{Color_eff} Y.L.\ Dokshitzer,  V.A.\ Khoze and S.I.\ Troyan, in ``Physics in Collision VI'',
Proceedings of the International Conference, Chicago, Illinois, 1986, ed. M.\ Derrick (World Scientific,
Singapore, 1987), p.\ 417;
J.D.\ Bjorken, Phys. Rev. D47, 101 (1993);
A.D.\ Martin, M.G.\ Ryskin and V.A.\ Khoze, Phys. Rev. D56, 5867, (1997);
O.J.P.\  Eboli, E.M.\ Gregores and  F.\ Halzen, hep-ph/9708283;
R.\ Oeckl, D.\ Zeppenfeld, hep-ph/9801257.  

\bibitem{CLS} H.\ Contopanagos, E.\ Laenen and G.\ Sterman,
Nucl.\ Phys.\ B484, 303 (1997).

\bibitem{KM} G.P.\ Korchemsky and G.\ Marchesini, Phys.\ Lett.\ B313, 433 (1993).

\bibitem{Brandtetal} R.A.\ Brandt, F.\ Neri, and M.-a.\ Sato, Phys. Rev. D24, 879 (1981).

\bibitem{eikrenorm1} A.M.\ Polyakov, Nucl. Phys. B164, 171 (1979);
I.Ya.\ Aref'eva, Phys. Lett. 93B, 347 (1980);
V.S.\ Dotsenko and S.N.\ Vergeles, Nucl. Phys. B169, 527 (1980).

\bibitem{KR} S.V.\ Ivanov and G.P.\ Korchemsky, Phys.\ Lett.\ B154, 197 (1985);
G.P.\ Korchemsky and A.V.\ Radyushkin, Phys.\ Lett.\ B171, 459 (1986);
Nucl.\ Phys.\ B283, 342 (1987); G.P.\ Korchemsky, Phys.\ Lett.\ B217, 330 (1989);
{\it ibid.}\ B220, 629 (1989).

\bibitem{Macfar} A.J.\ Macfarlane, A.\ Sudbery and P.H.\ Weisz,
Commun.\ Math.\ Phys. 11, 77 (1968).

\bibitem{Dixon} L.\ Dixon in {\it QCD and Beyond, Proceedings of the
Theoretical Advanced Study Institute in Elementary Particle Physics (TASI-95)},
ed.\ D.E.\ Soper (World Scientific, Singapore, 1996), p.\ 539,
hep-ph/9601359.

\bibitem{Bartels}  J.\ Bartels, Z.\ Phys.\ C60, 471 (1993). 


\end{thebibliography}
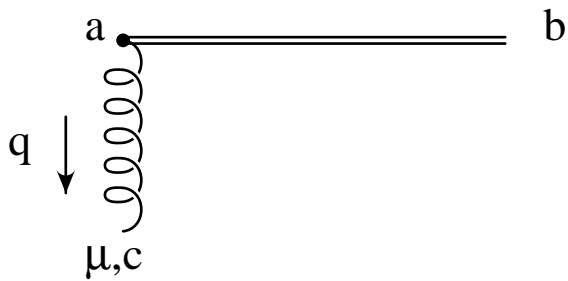
\end{document}